\documentclass[aps,superscriptaddress,amsmath,amssymb,twocolumn,amsfonts,floatfix,nofootinbib,longbibliography]{revtex4-1}
\usepackage{graphicx}
\usepackage{color}
\graphicspath{{Pictures/}}
\usepackage{hyperref}
\usepackage{soul}
\usepackage{natbib}
\hypersetup{pdfnewwindow=true, colorlinks=true, linkcolor=ForestGreen, anchorcolor=blue, citecolor=Purple, filecolor=blue, menucolor=blue, urlcolor=magenta}
\usepackage{cleveref}
\usepackage{color}
\usepackage{float}
\usepackage{soul}
\usepackage{braket}
\usepackage{enumitem}
\usepackage{marginnote}
\usepackage[normalem]{ulem}

\newcommand{\ie}{{\it i.e.,\,\,}}

\usepackage{amsmath}
\usepackage{mathtools}
\usepackage{braket}
\usepackage{graphicx}
\usepackage{amssymb}
\usepackage[dvipsnames]{xcolor}
\usepackage{multirow}
\usepackage{xfrac}
\usepackage{textcomp}
\usepackage{float}
\usepackage{enumerate}
\usepackage{subfigure}
\usepackage{stackengine}
\usepackage{booktabs}
\usepackage{physics}

\newcommand{\vect}[1]{\boldsymbol{\mathrm{#1}}}
\mathchardef\mhyphen="2D 

\newcommand\bea{\begin{eqnarray}}
	\newcommand\eea{\end{eqnarray}}
\newcommand\beq{\begin{equation}}  
	\newcommand\eeq{\end{equation}}

\newcommand{\non}{\nonumber}  

\usepackage{sidecap,tikz}

\begin{document}
	
	\title{Topological Majorana zero modes and the superconducting diode effect driven by Fulde-Ferrell-Larkin-Ovchinnikov pairing in a helical Shiba chain}
	
	\author{Sayak Bhowmik}
	\email{sayak.bhowmik@iopb.res.in}
	\affiliation{Institute of Physics, Sachivalaya Marg, Bhubaneswar, Orissa 751005, India}
	\affiliation{Homi Bhabha National Institute, Training School Complex, Anushakti Nagar, Mumbai 400094, India}
	
	\author{Arijit Saha}
	\email{arijit@iopb.res.in}
	\affiliation{Institute of Physics, Sachivalaya Marg, Bhubaneswar, Orissa 751005, India}
	\affiliation{Homi Bhabha National Institute, Training School Complex, Anushakti Nagar, Mumbai 400094, India}

	\begin{abstract}
		We propose a theoretical framework for the realization of Fulde-Ferrell-Larkin-Ovchinnikov (FFLO) pairing in a helical Shiba chain subjected to an out-of-plane Zeeman field, analyzed through a self-consistent Bogoliubov-de-Gennes (BdG) mean-field formalism approach. A chain of magnetic adatoms with helical spin texture deposited on the surface of a common $s$-wave superconductor, has emerged as a pivotal platform for realizing topological Majorana zero modes (MZMs). Our study reveals the crucial role of finite momentum pairing of Cooper pairs in the form of FFLO state which also supports topological MZMs at the ends of the chain. Interestingly, we demonstrate that FFLO pairing facilitates non-reciprocal charge transport, giving rise to superconducting diode effect in our system where both time-reversal and inversion symmetries are broken. Such diode effect stems directly from the presence of finite Cooper pair momentum of the FFLO ground state. Our comprehensive analysis highlights the intricate interplay between the richness of helical Shiba chain, the out-of-plane Zeeman field, and FFLO pairing in the emergence of MZMs and driving the superconducting diode effect. These findings offer valuable insights into the design and realization of topological superconducting devices with diode-like properties, potentially advancing technological applications in quantum computing and superconducting electronics.
	\end{abstract}
	
	\maketitle
	
	{\textcolor{blue}{\textit{Introduction}}}- In recent times, topological superconductors have gained immense attention  in modern quantum condensed matter research, as they offer elegant platform for hosting Majorana Zero modes (MZMs)-emergent zero-energy quasiparticle excitations 
at the boundaries~\cite{Kitaev_2001,Leijnse_2012,Zhang_Qi,Alicea_2012,Yuval_Oreg_Oppen, Lutchyn_Sau,Beenakker}.  These MZMs, renowned for their non-Abelian braiding statistics, has been  proposed  to form the backbone of topological quantum computation~\cite{Ivanov,KITAEV20032,Stern2010,CNayak}.  Following the foundational theoretical framework of  Kitaev model based on one dimensional (1D) $p$-wave superconductor~\cite{Kitaev_2001,Kitaev2009}, several setups including experimentally realizable Rashba nanowires proximitized to $s$-wave superconductors with applied Zeeman field, has been proposed to engineer Kitaev-like $p$-wave  physics~\cite{Leijnse_2012,Aguado2017,Alicea_2012,Lutchyn_Sau,Mourik}.  
In recent years, helical Shiba chains, formed by magnetic adatoms fabricated on $s$-wave superconducting substrates, emerge as an intriguing alternative setup for exploring topological superconductivity~\cite{Bernevig, Felix, LossRKKY,Teemu2d, LossU1}. The interplay of the magnetic texture and superconductivity give rise to Yu-Shiba-Rusinov (YSR) states, which hybridize to form Shiba bands within the parent superconducting gap. These Shiba bands can exhibit robust topological superconducting phases anchoring MZMs. They are protected by the minigap realized in various experimental setups featuring scanning tunneling microscopy (STM) measurements~\cite{SNadjexp, KimSciadv,Schneidernat,Crawford2022,Schneidernat2,Beck2021}. 
	
Previous theoretical and experimental studies in Shiba chain have predominantly investigated the conventional Bardeen Cooper Schrieffer (BCS)-type superconducting pairing, characterized by zero center-of-mass momentum of the Cooper pairs. However, Cooper pairs can acquire finite center-of-mass momentum $\boldsymbol{q}$, described by Fulde-Ferrell-Larkin-Ovchinnikov (FFLO) pairing~\cite{Fulde_1964,Larkin_1964} (involving  electrons with momentum $\boldsymbol{k+q/2}$ and 
$\boldsymbol{-k+q/2}$), driven by a shifted Fermi surface in presence of a Zeeman field(s) and Rashba spin-orbit coupling (SOC)~\cite{LiangPNAS,SLlicprl,Yanasediode,Picoli,Nagaosanjp,Qu2013natcom}. This intriguing phenomena of FFLO pairing stands as a key ingredient for realizing non-reciprocal charge transport, thus giving rise to the exotic physics of superconducting diode effect (SDE)~\cite{Nadeem2023,LiangPNAS,Lossdiode,Yanasediode,Yanaseprl,Picoli,Nagaosanjp}. The latter turns out to be a rapidly emerging topic with promising implications for quantum technologies and advancements.
	
In this letter, we extensively demonstrate the FFLO pairing mechanism in the 1D helical spin chain of Shiba lattice in presence of external Zeeman field and illustrate the impact of non-conventional 
finite-momentum pairing on the emergence of topologically protected MZMs. Based on our BdG mean-field self-consistent analysis, we reveal that the system undergoes a topological phase transition driven by the finite Cooper pair momentum $q$, exhibiting MZM signatures. While similar topological features are reported in a two-dimensional setup featuring FFLO superconductivity in spin-orbit coupled degenerate Fermi gases and external magnetic field~\cite{Qu2013natcom}; we present an alternative setup that hosts FFLO pairing assisted MZMs, uniquely emerging out of the intricate interplay between the spin spiral (SS), Zeeman field and superconductivity. Furthermore, we also illustrate the non-reciprocal nature of the critical currents, a direct consequence of the presence of finite momentum of Cooper pairs in the FFLO ground state or equilibrium state ($q_0$). This intriguing signatures establish our system as a promising platform for realizing  SDE.  We also discuss the possibility to eliminate the necessity of the external Zeeman field by choosing a particular 
type of SS to demonstrate the above mentioned features.

\clearpage
	\begin{figure}[t!]
		\centering \includegraphics[width=\columnwidth]{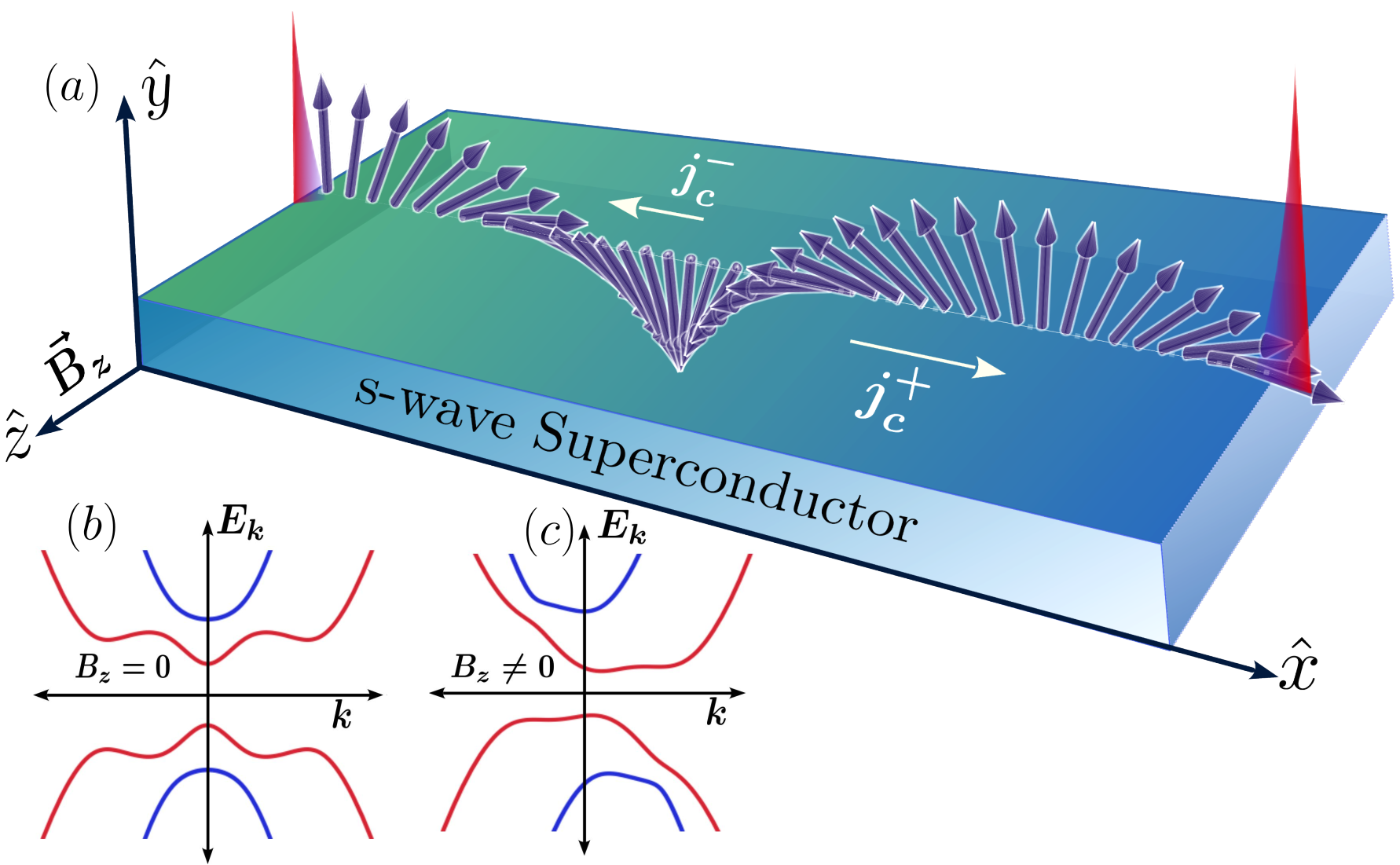}
		\caption{\textbf{Schematic of our setup:} (a) A 1D array of magnetic adatoms with spatially varying spin texture (purple arrows), confined to $xy$ plane (out of plane N\'{e}el type SS), are deposited on top of a common bulk $s$-wave superconductor in presence of an external Zeeman field $B_z$. In panels (b) and (c), Bogoliubov quasiparticle dispersions for this setup is depicted
in the absence and presence of the external field $B_z$ respectively.}
		\label{fig:Fig1}
	\end{figure}

{\textcolor{blue}{\textit{Model hamiltonian and mechanism of FFLO pairing}}}- The system under consideration features a SS texture (out-of-plane N\'{e}el type), comprising of a chain of spatially varying classical magnetic impurity atoms, integrated on the surface of a bulk $s$-wave superconductor (SC) with an external Zeeman field applied perpendicular to the plane of SS  
[see Fig.~\ref{fig:Fig1}(a)]. To realize finite momentum $s$-wave gap (FFLO pairing) in our proposed setup, we consider a local attractive Hubbard interaction $\mathcal{H}_I=-\frac{U}{2}\sum_{s,s^\prime} \int d^{3} {\bf{r}} c_{s}^\dagger({\bf{r}})c_{s^\prime}^\dagger({\bf{r}})c_{s^\prime}({\bf{r}})c_{s}({\bf{r}})$ within the bulk $s$-wave SC. Here, $U$ denotes the attractive interaction strength and $c_{s}({\bf{r}})$ annihilates an electron at position $\bf{r}$ with spin $s\in\{\uparrow,\downarrow\}$. Following the framework of mean-field approximation~\cite{Qu2013natcom,Yanasediode,Yanaseprl,sayakarxiv} we decouple $\mathcal{H}_I$ in the $s$-wave FFLO pairing channel, and further we consider these superconducting correlations to be proximity induced in the 1D spin chain~\cite{sayakarxiv}. Thus our self-consistent mean field treatment mimics intrinsically superconducting 1D chain with helical spin texture. Notably, the superconducting correlations ``leak" from the bulk superconductor to the Shiba chain, and the magnitude of the proximity-induced gap depends on the details of the system (quality of the interface, 
SS, magnetic field etc). In general, these ``leaked" correlations in the 1D Shiba lattice are not constrained by the same self-consistency condition as in intrinsically superconducting systems. However, 
as a zeroth order approximation, we assume that the self-consistency condition applies similarly, apart from renormalization of the pairing gap. This assumption is justified by the fact that the proximity-induced pairing originates from a high-symmetry three-dimensional superconductor, which effectively suppresses fluctuations~\cite{Picoli,sayakarxiv}.
Hence, the Bogoliubov-de-Gennes (BdG) mean-field Hamiltonian for the low-energy continuum model represented in the workable Nambu basis: $\Psi(x)=[c_{\uparrow}(x),c_{\downarrow}(x),c_{\downarrow}^\dagger(x),-c_{\uparrow}^\dagger(x)]^T$, can be 
obtained as
	\bea
	\mathcal{H}  =  \frac{1}{2}\int dx \, \Psi^\dagger(x) \mathcal{H}_{BdG}(x) \Psi(x) + \frac{L}{U} |\Delta|^2  \ , \non \\
	\mathcal{H}_{BdG} (x)   =    
	\begin{bmatrix}
		\hat{h}(x) & \hat{\Delta}(x) \sigma_0 \\
		\hat{\Delta}^*(x) \sigma_0 & -\sigma_y \hat{h}^*(x) \sigma_y
	\end{bmatrix} \ , \non \\
	\hat{h}(x) = \left(-\frac{\hbar^2}{2m} \nabla_{x}^2 - \mu\right)\sigma_{0} - J \, \vect{S}({x}) \cdot \boldsymbol{\sigma} + B_z \sigma_z \ ,
	\label{eqn:1}
	\eea
	where, $L$ being the length of the magnetic chain and ${\sigma}$ denotes the Pauli matrices in spin space. Here $\Delta(x)= \Delta  e^{iqx}$ represents the $s$-wave FFLO order parameter, where $q$ symbolizes the Cooper pair momentum and $J,\mu$ refer to the local exchange coupling strength between the  electron spin of SC and the magnetic impurity atoms, chemical potential respectively. The local spin vector of the impurity atoms: $\vect{S}(x)$=$\lvert \vect{S} \rvert\begin{pmatrix} \cos[\phi(x)],\!&\! \sin[\phi(x)],\! &\! 0 \end{pmatrix}$ \cite{SStexture_ANandy,Pritamprb1, LossU1} is considered to be classical and restricted (not necessarily) within the $xy$ plane with $|\vect{S}|=1$. Also, $\phi(x)$=$\left(g x\right)$ represents the spatial variation of the SS texture (out-of-plane N\'{e}el type~\cite{SStexture_ANandy,Pritamprb1}), where $g$ is the angle between two adjacent spins.   
	
Subsequently, we perform an unitary transformation: ${U^{\prime}}^\dagger \mathcal{H}_{BdG} (x) U^{\prime}$ with  $ U^{\prime}=e^{-\frac{i}{2}\phi(\vect{r})\sigma_z}$~\cite{LossU1,Pritamprbl,Pritamprb}, followed by a Fourier transformation and the momentum space Hamiltonian is readily obtained as: $\mathcal{H} = \frac{1}{2} \sum_{k}\Psi_{k}^\dagger\mathcal{H}_{BdG} (k) \Psi_{k}+ \frac{L}{U}|\Delta|^2 $, where  
	\bea
	\mathcal{H}_{BDG}(k)=
	\begin{bmatrix}
		\hat{h}(k + \frac{q}{2}) &  \Delta \sigma_0 \\
		\Delta \sigma_0 & -\sigma_y \hat{h}(-k+\frac{q}{2})^*\sigma_y
	\end{bmatrix}\ ,\non \\
	\hat{h}_k = \xi_{{k},{g}} + \frac{1}{2}{g}{k}\sigma_z + J\sigma_x +B_z\sigma_z \ .
	\label{eqn:2}
	\eea
	Here, $\xi_{{k},{g}}= [k^2/2-(\mu-g^2/2)]\sigma_0$  and the Nambu basis reads as $\Psi({k})$=$[c_{{k+q/2},\uparrow}, c_{{k+q/2},\downarrow}, c^{\dagger}_{{-k+q/2},\downarrow}, -c^{\dagger}_{{-k+q/2},\uparrow}] ^{\rm T}\!$. We consider $(\hbar,m)=1$, throughout our analysis for simplicity. Intriguingly, an effective intrinsic SOC term: $\frac{1}{2} {g}{k}~\sigma_z$  and Zeeman term: $J\sigma_x$ emerge out of the considered SS leading to a rich platform for exploring topological MZMs
and SDE. However, the finite Cooper pair momentum: $q $ is realized via the externally applied Zeeman field $B_z$ that lies out of the plane of spin texture. Note that, $B_z$ introduces asymmetry in the BDG bands as shown in Fig.~\ref{fig:Fig1}(b),(c), thus favoring $s$-wave FFLO superconducting state that can host topological MZMs, assisted by $q$. 
	\begin{figure}[b]
		\centering \includegraphics[width=1.0\columnwidth]{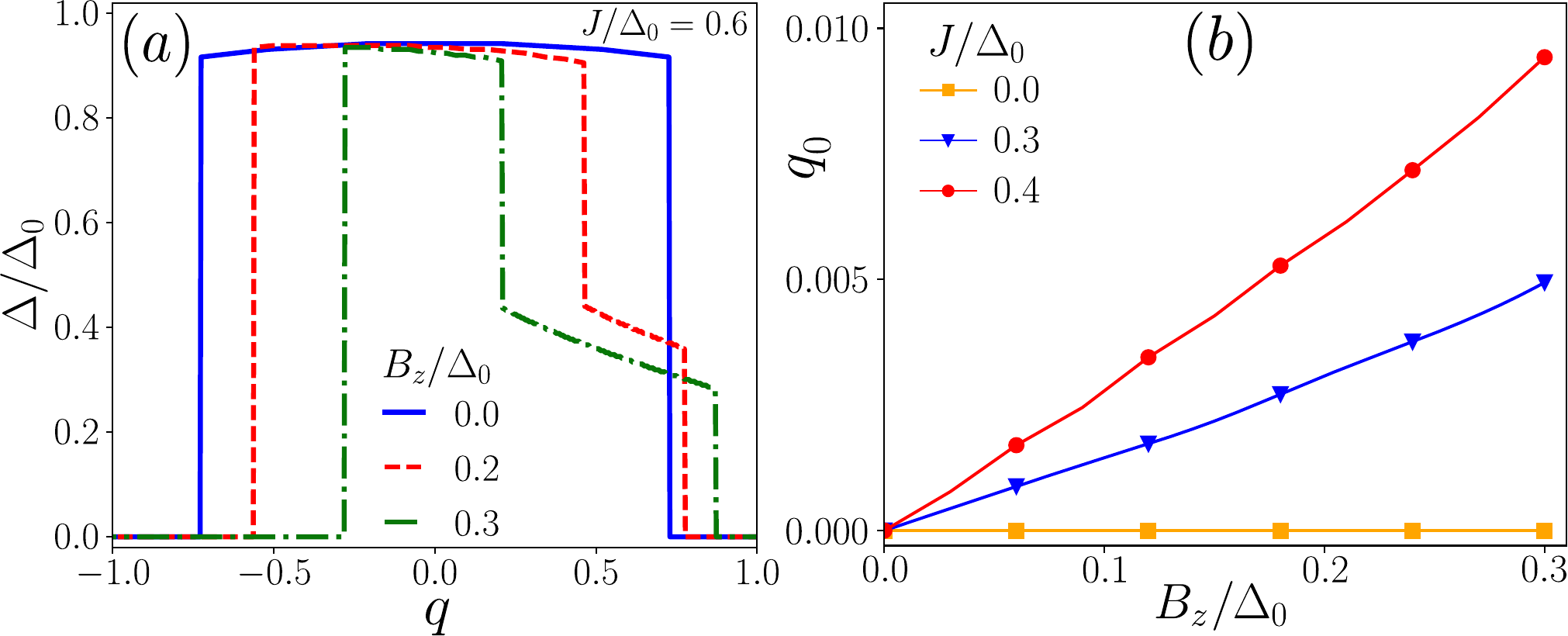}
		\caption{\textbf{Stability of FFLO pairing and  ground state:} In panel (a) we show the self-consistent superconducting gap $\Delta(q)$ as a function of $q$ choosing different values of external field $B_z$  with  $J/\Delta_0=0.6$. The $\Delta(q)$ profile is symmetric with respect to $q$ when $B_z=0$. However, presence of non zero  $B_z$ results in asymmetry of $\Delta(q)$ profile. Panel 
(b) demonstrates the FFLO ground state Cooper pair momentum $q_0$ as a function of the external field $B_z$ for various values of $J$, clearly indicating that the presence of non zero $B_z$ and 
$J$ support FFLO pairing in our superconducting setup featuring helical spin texture. The other system parameters are chosen as $(\mu/\Delta_0, \beta^{-1}, U)=(1, 0.01 \text{meV}, 0.358 \text{meV})$.} \label{fig:Fig2}
	\end{figure}
	To describe the mechanism of finite momentum pairing we employ a self-consistent BdG 
mean field approach. The FFLO superconduting order parameter $\Delta(q)$ is self-consistently determined by minimizing the condensation energy: $\Omega(q,\Delta) = F(q,\Delta) - F(q,0)$ for a particular value of $q$.  Here, $F(q,\Delta)$ denotes the free energy density of the Hamiltonian in Eq.~(\ref{eqn:2}), given as
	\beq
	F(q,\Delta) = -\frac{1}{L\beta}\sum_{i,k} \ln\left[1+e^{-\beta E_{i,k}(q)}\right] + \frac{|\Delta(q)|^2}{U}\ ,
	\label{eqn:3}
	\eeq
	where, $\beta=(k_B T)^{-1}$ with $T$ being the temperature and $k_B$ the Boltzmann constant and $i$ denotes the BdG band index. Having obtained $\Delta(q)$ \ie~the $s$-wave FFLO stability [see Fig~\ref{fig:Fig2}(a)], we further determine the FFLO superconducting ground state Cooper pair momentum: $q_0$ by optimizing $\Omega(q,\Delta)$ with respect to $q$ by solving 
	\beq
	\frac{\delta \Omega(q,\Delta) }{\delta q}\bigg|_{q=q_0}=0\ .
	\label{eqn:ground}
	\eeq       
	As displayed in Fig.~\ref{fig:Fig2}(b), the system exhibits a finite $q_0$ only when a non-zero $B_z$ is present with a linear relationship between them,  thus highlighting the pivotal role of $B_z$ in driving the FFLO pairing mechanism in our setup.  The BCS gap without the presence the spin texture and external field: $(B_z,J=0)$, is denoted by $\Delta_0$.   We self-consistently set $\Delta_0=1$ meV for the entire analysis in this article.  For a give set of system parameters, the superconducting regime is defined within the interval  $q^- \leq q \leq q^+$, beyond which the superconducting gap vanishes. Furthermore, the presence of a finite $B_z$ introduces asymmetry in $\Delta(q)$: $\Delta(q)\ne\Delta(-q)$ as depicted in Fig.~\ref{fig:Fig2}(a). This asymmetry results in non-reciprocal charge transport in the system, thereby supporting  SDE (see latter text for discussion).
	
	
	{\textcolor{blue}{\textit{Signatures of toplogically protected MZMs in FFLO superconducting phase}}}- 
	In order to demonstrate the topological features of our proposed setup and capture the MZM signatures, we systematically proceed by considering a minimal Hamiltonian realized in a 1D lattice 
	subject to open boundary condition (OBC):   
	\begin{eqnarray}
		H_l&=& \! - t \!\sum_{\langle m,n \rangle,s}   c_{m,s}^{\dagger}  c_{n,s} \! + \sum_{n} \Delta e^{iqn}
		(c_{n,\uparrow}^{\dagger}c_{n,\downarrow}^{\dagger}\!+\! {\rm H.c.})  \nonumber\\
		&&+J  \! \sum_{n,s,{s}^\prime} \! \! c_{n,s}^{\dagger} \! \left(\vect{S}_n \! \cdot \! \vect{\sigma}\right)_{s, {s}^\prime} c_{n,{s}^\prime} -\mu \sum_{n,s} c_{n,s}^{\dagger} c_{n,s} \! \! \nonumber\\ 
		&&+ B_z  \! \sum_{n,s,{s}^\prime} \! \! c_{n,s}^{\dagger} (\sigma_z)_{s, {s}^\prime} c_{n,{s}^\prime}\ . \quad ~
		\label{lattice}
	\end{eqnarray}
	where, $t$ represents the hopping amplitude between nearest neighbor sites $\langle m,n \rangle$, and $S_n = \begin{pmatrix} \cos[\phi(n)], & \sin[\phi(n)], & 0 \end{pmatrix}$ denotes the local  
	spin vector at site $n$, providing a realization of the SS texture in the lattice, where $\phi(n+1) - \phi(n) = g$. The remaining parameters denote the same as defined after Eq.~(\ref{eqn:1}). 
	
\begin{figure}[t]
		\centering \includegraphics[width=1.0\linewidth]{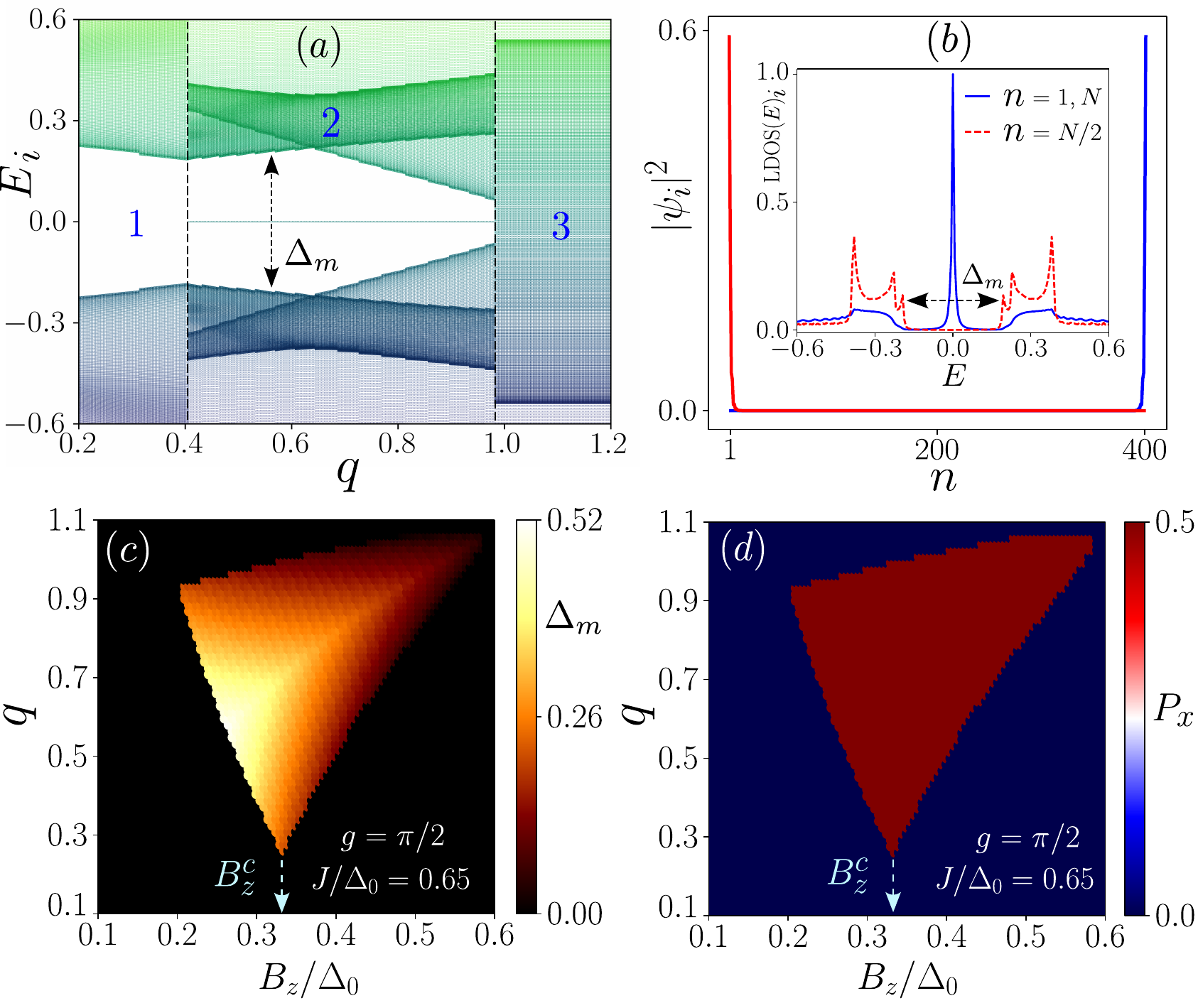}
		\caption{\textbf{Topological signatures of MZMs:} Numerical results are presented for a finite lattice of 400 sites obeying OBC applied to the BDG lattice Hamiltonian [Eq.~(\ref{lattice})]. 
		The  eigenvalue spectrum as a function of Cooper pair momentum is shown in panel (a) incorporating the self consistent solution $\Delta(q)$ corresponding to the model parameters ($B_z/
		\Delta_0, J/\Delta_0, \mu/\Delta_0, g,t$) =($0.3, 0.65, 1.0, \pi/2, 0.5$). As indicated in the spectrum, there are three distinct regions (1): trivially gapped, (2): topologically gapped region hosting 
		MZMs protected by the effective minigap $\Delta_m$ and (3): normal region where $\Delta(q)=0$. (b) The corresponding site resolved normalized probability density $|\psi_i|^2 $ of the 
		two MZMs is shown at $q=0.7$ exhibiting strong localization of them at the end of the chain. In the inset, the energy resolved LDOS$(E)$ is depicted at the end (middle) of the chain indicated 
		by solid line (dashed red line). The effective minigap $\Delta_m$ profile and the topological invariant: $P_x$ (bulk dipole moment) is visually showcased in the $B_z - q$ plane in panels (c) 
		and (d), respectively. Other model parameters are chosen as $(U, \beta^{-1})=(2.78, 0.01)~\text{meV}$.}
		\label{fig:Fig3}
	\end{figure}

	Afterwards, we determine the $s$-wave FFLO order parameter $\Delta$ following a similar self-consistent formalism described earlier, but now applied to the real-space model defined by the lattice Hamiltonian in Eq.~(\ref{lattice}) (see supplementary material (SM) for detailed analysis~\cite{supp}). Note that, the BCS gap denoted by $\Delta_0$, is set to $1$~meV for the considered lattice. By utilizing the self-consistent solution of $\Delta$, the BdG quasi-particle energy eigenvalue spectrum is obtained by diagonalizing the lattice Hamiltonian employing open boundary condition (OBC). As illustrated in Fig.~\ref{fig:Fig3}(a), the BdG spectrum reveals three distinct phases across different regimes when depicted as a function of Cooper pair momentum $q$. Regime~(1) depicts trivially gapped phase.
However, in-gap states at zero energy (doubley degenerate) within the BdG quasi particle gap appears in Regime (2). This indicates the topological phase featuring MZMs, driven by finite $q$. 
The physical reason behind the emergence of MZMs, supported by FFLO state can be attributed to the fact that the emergent $p$-wave pairing still remains present in the background of FFLO state within the topological regime (dependent on the system parameters $\mu, J,B_z,q,\Delta(q)$), leading to a mixed superconducting state with both $s$-wave FFLO and $p$-wave components
(in the form of effective minigap $\Delta_m$). This interplay allows the system to sustain topological superconductivity hosting MZMs even in the presence of FFLO order.
	On the other hand, Regime (3) represents gapless normal phase, signifying the absence of superconductivity $\Delta=0$. 
\begin{figure*}[htb!]
	\centering
	\includegraphics[width=1.0\linewidth]{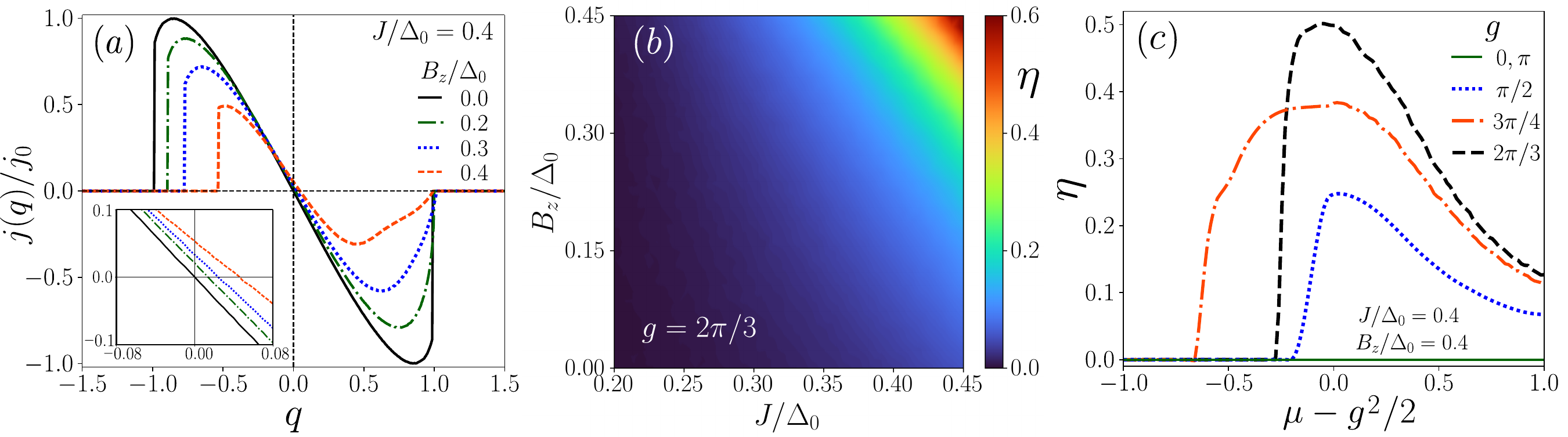}
		\caption{\textbf{Superconducting diode effect:} The supercurrent density ($j$) is displayed as a function of $q$ for different values of $B_z$ with $J/\Delta_0=4$ in panel (a), highlighting the nonreciprocal nature of the critical currents: $|J_{c}^+|\ne|J_{c}^-|$ when $B_z\neq 0$. Here, $j_0 \equiv j_c ( B_z = 0, J/\Delta_0=0.4)$. The shift in equilibrium state (FFLO) Cooper pair momentum for which $j(q_0)=0$ is depicted in the inset. Panel (b) illustrates the diode efficiency ($\eta$) in the $J - B_z$ plane choosing $g=2\pi/3$. The dependence of $\eta$ on the chemical potential $\mu$ is shown in panel (c) for different choices of $g$ with ($J/\Delta_0, B_z/\Delta_0=0.4$), manifesting maximum efficiency at $\mu \approx g^2/2$. The other system parameters are chosen as: 
		$(U, \beta^{-1})$=($0.358$ \text{meV}, $0.1$ \text{meV}).}
	\label{fig:Fig4}
\end{figure*}	
	
	As visually depicted through the site resolved normalized probability distribution: $|\psi_i|^2$ in Fig.~\ref{fig:Fig3}(b), the two MZMs exhibit sharp localization at the two ends of the chain. In the 
	topological regime, the energy resolved local density of states (LDOS)$(E)$ demonstrates the edge localization of the MZMs that is well protected by the  effective minigap $\Delta_m$. 
	In contrast, the LDOS$(E)$ at the middle of the chain just reflects the minigap and normal YSR states [see the inset of Fig.~\ref{fig:Fig3}(b)]. These features strongly underscores the signatures 
	of topological MZMs driven by finite Cooper pair momentun $q$. 
	
	{\textcolor{blue}{\textit{Topological characterization}}}-To accurately distinguish the topological phase from the trivially gapped phase, we employ the bulk dipole moment (polarization $P_x$) 
	as the suitable topological invariant~\cite{Resta,Wheeler,KangPRB}, defined as:
	\begin{equation}
		P_{x}= \frac{1}{2 \pi} {\rm Im} \left[ \Trace \left\{\ln \left(\mathcal{U}^\dagger \mathcal{W} \mathcal{U} \right)  \right\}\right] \ ,
		\label{dipole}
	\end{equation}
	where, $\mathcal{U}$ represents $N \!\times \! N_{\textrm {occ}}$ matrix constructed by columnwise stacking the occupied energy eigenstates of the BdG lattice Hamiltonian [Eq.~(\ref{lattice})]. The operator $\mathcal{W}$=$\exp \left[ i 2 \pi \hat{p}\right]$ with $\hat{p}=\hat{x}/L$ being the microscopic dipole operator defined for a system of size $L$, where $\hat{x}$ denotes the position operator. $P_x=0.5$ distinctly identifies the topological phase exhibiting MZM signatures, while $P_x=0$ indicates the trivially gapped phase.

	Additionally, in Fig.~\ref{fig:Fig3}(c) we demonstrate the minigap $\Delta_m$ spectrum in the $(B_z$-$ q)$ plane. The appearance of finite $\Delta_m$ within the Shiba band also serves as a reliable indicator of the topological phase with $p_x=0.5$ [see Figs.~\ref{fig:Fig3}(c),(d)], thereby justifying the topological protection of MZMs by the minigap. For a given set of system parameters there exists a critical value of the external field $B^{c}_z$ above which $\Delta(q=0)=0$, \ie the system supports finite $\Delta$ only for $q>0$ signifying true FFLO state. Consequently, the following is also reflected 
	in the characteristic nature of the minigap spectrum as well as $P_x$ as depicted in Figs.~\ref{fig:Fig3}(c),(d). It is important to note that all the results presented above were computed implementing the self-consistent solution for $\Delta(q)$ in presence of all relevant parameters ($B_z, J, T, \mu, t$). While smooth gap-closing transitions in the BdG spectrum as a function of continuously varying parameters are often depicted in studies of band topology, these typically assume a constant $\Delta$ throughout the parameter range. However, for completeness we address these issues 
	in the SM~\cite{supp}.                     
	
	{\textcolor{blue}{\textit{Realizing non-reciprocal charge transport (SDE)}} - Having established the FFLO pairing mechanism in our proposed setup and the topological MZM signatures, we next seek to elucidate the intriguing phenomenon of SDE which is a characteristic signature of FFLO pairing. Superconducting systems with broken time-reversal and inversion symmetries often serve as favourable platforms for exploring SDE~\cite{Nadeem2023,LiangPNAS,Lossdiode,Yanasediode,Yanaseprl,Picoli,Nagaosanjp}. The presence of the SS readily breaks both the inversion symmetry: $ J \vect{S}({x}) \cdot\boldsymbol{\sigma} \ne \pm J \vect{S}({-x}) \cdot\boldsymbol{\sigma}$ and time reversal symmetry: $\mathcal{T}^{-1} J \vect{S}({x}) \cdot\boldsymbol{\sigma} \mathcal{T} \ne J \vect{S}({x}) \cdot\boldsymbol{\sigma}$, where $\mathcal{T}= i \sigma_{y} \mathcal{K}$ is the time reversal operator with $\mathcal{K}$ being the complex conjugation operator. Inversion symmetry is preserved for the trivial cases when $g=0 (\pi)$ where the spin texture represents ferromagnet (anti-ferromagnet). We explicitly show that these regular configurations of the SS does not support FFLO pairing when realized in a lattice, even in presence of $B_z$ (see SM for details~\cite{supp}).   
		
	SDE can be realized via the presence of unequal critical super currents in opposite directions: $j^{+}_c\ne-j^{-}_c $. We extensively demonstrate that the non reciprocity of the critical current is a direct consequence of the presence of finite $q_0$ (\ie finite $B_z$) in our system. The supercurrent density is computed form the condensation energy $\Omega(q,\Delta)$ in the following way~\cite{LiangPNAS,Lossdiode}:
		\begin{equation}
			j(q)=-2e\,\frac{\partial\Omega(q,\Delta)}{\partial q}\ ,
			\label{current}
		\end{equation}		
	by incorporating the self-consistent solutions of $\Delta(q)$. Here $e$ denotes the electronic charge and the sign of $j$ reveals the flow direction. From Eq.~(\ref{eqn:ground},\ref{current}) it is evident that $j(q_0)=0$.  
		
	The behaviour of $j(q)$  is visually depicted in Fig.~\ref{fig:Fig4}(a) for different values of external field $B_z$.  In the absence of $B_z$, super-current is symmetric with respect to $q$ \ie 
	$ j(q) = -j(-q) $. However, when $B_z \neq 0$, the behavior of $j(q)$ exhibits asymmetry: $ j(q) \ne -j(-q) $, thereby highlighting non-reciprocal behavior. This non-reciprocity is a direct 
	consequence of the fact that, the FFLO ground state Cooper pair momentum: $q_0$ for which $j(q_0)=0$, becomes non-zero when $B_z \neq 0$ [see the inset of Fig.~\ref{fig:Fig4}(a)]. 
		
	Within the superconducting region $q^- \leq q \leq q^+ $ there exist maximal super current in both the flow directions $(j^{+}_c, j^{-}_c)$, also known as critical currents, exceeding which leads to 
	breakdown of superconductivity. In presence of $B_z$ these critical currents are non reciprocal: $j^{+}_c \ne  -j^{-}_c$, as shown in Fig.~\ref{fig:Fig4}(a). This non-reciprocity of the critical currents 
	gives rise to the remarkable phenomena of SDE in our system, characterized by the diode efficiency $\eta$, defined as~\cite{LiangPNAS}  
		\begin{equation}
			\eta=\frac{|J_{c}^+|-|J_{c}^-|}{|J_{c}^+|+|J_{c}^-|}\ .
			\label{eff}
		\end{equation}      
	The variation  of $\eta$ with respect to parameters $B_z,J,\mu,g $ is visually depicted in Figs.~\ref{fig:Fig4}(b),(c). Intriguingly, our system exhibits a notably high diode efficiency of approximately $60\%$ [see Fig.~\ref{fig:Fig4}(b)] for an optimal set of system parameters. As shown in Fig.~\ref{fig:Fig4}(c), the diode efficiency $\eta$ peaks around $\mu\approx g^2/2$, where the re-normalized chemical potential vanishes. This characteristic feature of superconducting diodes is also reported in preceding investigations availing other systems~\cite{Lossdiode,Nagaosanjp}. As expected from symmetry arguments, our system does not demonstrate SDE for the trivial choice of the spin texture when $g=0,\pi$ as shown in Fig.~\ref{fig:Fig4}(c).

	{\textcolor{blue}{\textit{Summary and Discussions}}} \--- In this article, we systematically demonstrate the emergence of asymmetric FFLO pairing mechanism in our proposed setup, which features 
	a out-of-plane N\'{e}el type SS fabricated on the surface of a common $s$-wave superconductor. Following the BdG mean-field approach, we present our results of topological superconducting 
	phase hosting MZMs and SDE driven by FFLO pairing via a self-consistent analysis by considering both a continuum model (momentum space) and a finite-size lattice model (real space). 
	Note that, in our analysis, the effect of topology on the SDE is not addressed, as we don't consider any contributions in that arising from the MZMs localized at the two ends of the 1D Shiba lattice.
		
	It is important to note that, the superconducting state with a Cooper pair momentum $q$ is accessed through tuning externally injected current also leading to reduction of superconducting gap 
	$\Delta(q)$~\cite{LiangPNAS,Picoli}. On the other hand, the presence of the external field $B_z$ supports the intrinsic mechanism of FFLO pairing in the system, shifting the Cooper pair momentum of the equilibrium state (or FFLO ground state) by a value $q_0$, ultimately leading to non reciprocity of $j(q)$. However, superconductivity and external magnetic field always manifest detrimental effect. Moreover, the presence of external $B_z$ could possibly lead to precession effects on the array of magnetic moments forming the Shiba lattice, potentially causing the system to become dynamical. 
	We assume such precession effects to be negligibly small for the classical spins due to their large total angular momentum. 
	Interestingly, it is possible to rule out the necessity of the external field $B_z$ by choosing a particular type of SS of the form: $\vect{{S}^{\prime}}(x)$=$\lvert \vect{S} \rvert\begin{pmatrix} \sin[\theta]\cos[\phi(x)],\!&\! \sin[\theta]\sin[\phi(x)],\! &\!\cos[\theta] \end{pmatrix}$ in Hamiltonian 
	[Eq.~(\ref{eqn:1})], where $\theta$ is fixed within the range $(0 < \theta < \pi/2) $. This leads to the following modification in Eq.~(\ref{eqn:1}) as $\hat{h}^{\prime}(x) = \left(-\frac{\hbar^2}{2m} \nabla_{x}^2 - \mu\right)\sigma_{0} - J^{\prime} \, \vect{S^{\prime}}({x}) \cdot \boldsymbol{\sigma}$. By following a similar approach as described above,  one can readily obtain $\mathcal{H}_{BdG}(k)$ 
	by replacing $\hat{h}_k$ in Eq.~(\ref{eqn:2}) by
		\begin{equation}
			\hat{h}^{\prime}_k = \xi_{{k},{g}} + \frac{1}{2} {g}{k}~  \sigma_z + J^{\prime}\sin[\theta]\sigma_x +J^{\prime}\cos[\theta]\sigma_z\ .
		\end{equation}  
	It is evident that, the effect of external $B_z$ can be readily realized through the effective Zeeman term $J^{\prime}\cos[\theta]\sigma_z$ that is intrinsically generated by the spin texture 
	$\vect{{S}^{\prime}}(x)$, hence providing a elegant platform to explore SDE in absence of any external fields. Without any loss of generality, all the above 1D results can be exactly obtained by 
	choosing the new parammeters ($J^{\prime}, \theta$) in place of $J, B_z$.

	{\textcolor{blue}{\textit{Acknowledgments}}} \---  S.B. and A.S. acknowledge the SAMKHYA: HPC Facility provided at IOP, Bhubaneswar and the two workstations provided by IOP, Bhubaneshwar from  DAE APEX project for the numerical computations.

		{\textcolor{blue}{\textit{Data Availibility Statement}}} \--- The datasets generated and analyzed during the current study are available from the corresponding author upon reasonable request.

\bibliography{ref}

\begin{thebibliography}{45}%
\makeatletter
\providecommand \@ifxundefined [1]{%
 \@ifx{#1\undefined}
}%
\providecommand \@ifnum [1]{%
 \ifnum #1\expandafter \@firstoftwo
 \else \expandafter \@secondoftwo
 \fi
}%
\providecommand \@ifx [1]{%
 \ifx #1\expandafter \@firstoftwo
 \else \expandafter \@secondoftwo
 \fi
}%
\providecommand \natexlab [1]{#1}%
\providecommand \enquote  [1]{``#1''}%
\providecommand \bibnamefont  [1]{#1}%
\providecommand \bibfnamefont [1]{#1}%
\providecommand \citenamefont [1]{#1}%
\providecommand \href@noop [0]{\@secondoftwo}%
\providecommand \href [0]{\begingroup \@sanitize@url \@href}%
\providecommand \@href[1]{\@@startlink{#1}\@@href}%
\providecommand \@@href[1]{\endgroup#1\@@endlink}%
\providecommand \@sanitize@url [0]{\catcode `\\12\catcode `\$12\catcode
  `\&12\catcode `\#12\catcode `\^12\catcode `\_12\catcode `\%12\relax}%
\providecommand \@@startlink[1]{}%
\providecommand \@@endlink[0]{}%
\providecommand \url  [0]{\begingroup\@sanitize@url \@url }%
\providecommand \@url [1]{\endgroup\@href {#1}{\urlprefix }}%
\providecommand \urlprefix  [0]{URL }%
\providecommand \Eprint [0]{\href }%
\providecommand \doibase [0]{http://dx.doi.org/}%
\providecommand \selectlanguage [0]{\@gobble}%
\providecommand \bibinfo  [0]{\@secondoftwo}%
\providecommand \bibfield  [0]{\@secondoftwo}%
\providecommand \translation [1]{[#1]}%
\providecommand \BibitemOpen [0]{}%
\providecommand \bibitemStop [0]{}%
\providecommand \bibitemNoStop [0]{.\EOS\space}%
\providecommand \EOS [0]{\spacefactor3000\relax}%
\providecommand \BibitemShut  [1]{\csname bibitem#1\endcsname}%
\let\auto@bib@innerbib\@empty
\bibitem [{\citenamefont {Kitaev}(2001)}]{Kitaev_2001}%
  \BibitemOpen
  \bibfield  {author} {\bibinfo {author} {\bibfnamefont {A~Yu}\ \bibnamefont
  {Kitaev}},\ }\bibfield  {title} {\enquote {\bibinfo {title} {Unpaired
  majorana fermions in quantumwires},}\ }\href {\doibase
  10.1070/1063-7869/44/10S/S29} {\bibfield  {journal} {\bibinfo  {journal}
  {Physics-Uspekhi}\ }\textbf {\bibinfo {volume} {44}},\ \bibinfo {pages} {131}
  (\bibinfo {year} {2001})}\BibitemShut {NoStop}%
\bibitem [{\citenamefont {Leijnse}\ and\ \citenamefont
  {Flensberg}(2012)}]{Leijnse_2012}%
  \BibitemOpen
  \bibfield  {author} {\bibinfo {author} {\bibfnamefont {Martin}\ \bibnamefont
  {Leijnse}}\ and\ \bibinfo {author} {\bibfnamefont {Karsten}\ \bibnamefont
  {Flensberg}},\ }\bibfield  {title} {\enquote {\bibinfo {title} {Introduction
  to topological superconductivity and majorana fermions},}\ }\href {\doibase
  10.1088/0268-1242/27/12/124003} {\bibfield  {journal} {\bibinfo  {journal}
  {Semiconductor Science and Technology}\ }\textbf {\bibinfo {volume} {27}},\
  \bibinfo {pages} {124003} (\bibinfo {year} {2012})}\BibitemShut {NoStop}%
\bibitem [{\citenamefont {Qi}\ and\ \citenamefont {Zhang}(2011)}]{Zhang_Qi}%
  \BibitemOpen
  \bibfield  {author} {\bibinfo {author} {\bibfnamefont {Xiao-Liang}\
  \bibnamefont {Qi}}\ and\ \bibinfo {author} {\bibfnamefont {Shou-Cheng}\
  \bibnamefont {Zhang}},\ }\bibfield  {title} {\enquote {\bibinfo {title}
  {Topological insulators and superconductors},}\ }\href {\doibase
  10.1103/RevModPhys.83.1057} {\bibfield  {journal} {\bibinfo  {journal} {Rev.
  Mod. Phys.}\ }\textbf {\bibinfo {volume} {83}},\ \bibinfo {pages}
  {1057--1110} (\bibinfo {year} {2011})}\BibitemShut {NoStop}%
\bibitem [{\citenamefont {Alicea}(2012)}]{Alicea_2012}%
  \BibitemOpen
  \bibfield  {author} {\bibinfo {author} {\bibfnamefont {Jason}\ \bibnamefont
  {Alicea}},\ }\bibfield  {title} {\enquote {\bibinfo {title} {New directions
  in the pursuit of majorana fermions in solid state systems},}\ }\href
  {\doibase 10.1088/0034-4885/75/7/076501} {\bibfield  {journal} {\bibinfo
  {journal} {Reports on Progress in Physics}\ }\textbf {\bibinfo {volume}
  {75}},\ \bibinfo {pages} {076501} (\bibinfo {year} {2012})}\BibitemShut
  {NoStop}%
\bibitem [{\citenamefont {Oreg}\ \emph {et~al.}(2010)\citenamefont {Oreg},
  \citenamefont {Refael},\ and\ \citenamefont {von Oppen}}]{Yuval_Oreg_Oppen}%
  \BibitemOpen
  \bibfield  {author} {\bibinfo {author} {\bibfnamefont {Yuval}\ \bibnamefont
  {Oreg}}, \bibinfo {author} {\bibfnamefont {Gil}\ \bibnamefont {Refael}}, \
  and\ \bibinfo {author} {\bibfnamefont {Felix}\ \bibnamefont {von Oppen}},\
  }\bibfield  {title} {\enquote {\bibinfo {title} {Helical liquids and majorana
  bound states in quantum wires},}\ }\href {\doibase
  10.1103/PhysRevLett.105.177002} {\bibfield  {journal} {\bibinfo  {journal}
  {Phys. Rev. Lett.}\ }\textbf {\bibinfo {volume} {105}},\ \bibinfo {pages}
  {177002} (\bibinfo {year} {2010})}\BibitemShut {NoStop}%
\bibitem [{\citenamefont {Lutchyn}\ \emph {et~al.}(2010)\citenamefont
  {Lutchyn}, \citenamefont {Sau},\ and\ \citenamefont
  {Das~Sarma}}]{Lutchyn_Sau}%
  \BibitemOpen
  \bibfield  {author} {\bibinfo {author} {\bibfnamefont {Roman~M.}\
  \bibnamefont {Lutchyn}}, \bibinfo {author} {\bibfnamefont {Jay~D.}\
  \bibnamefont {Sau}}, \ and\ \bibinfo {author} {\bibfnamefont
  {S.}~\bibnamefont {Das~Sarma}},\ }\bibfield  {title} {\enquote {\bibinfo
  {title} {Majorana fermions and a topological phase transition in
  semiconductor-superconductor heterostructures},}\ }\href {\doibase
  10.1103/PhysRevLett.105.077001} {\bibfield  {journal} {\bibinfo  {journal}
  {Phys. Rev. Lett.}\ }\textbf {\bibinfo {volume} {105}},\ \bibinfo {pages}
  {077001} (\bibinfo {year} {2010})}\BibitemShut {NoStop}%
\bibitem [{\citenamefont {Beenakker}(2013)}]{Beenakker}%
  \BibitemOpen
  \bibfield  {author} {\bibinfo {author} {\bibfnamefont {C.W.J.}\ \bibnamefont
  {Beenakker}},\ }\bibfield  {title} {\enquote {\bibinfo {title} {Majorana
  fermions in superconductors},}\ }\href {\doibase
  10.1146/annurev-conmatphys-030212-184337} {\bibfield  {journal} {\bibinfo
  {journal} {Annual Review of Condensed Matter Physics}\ }\textbf {\bibinfo
  {volume} {4}},\ \bibinfo {pages} {113--136} (\bibinfo {year}
  {2013})}\BibitemShut {NoStop}%
\bibitem [{\citenamefont {Ivanov}(2001)}]{Ivanov}%
  \BibitemOpen
  \bibfield  {author} {\bibinfo {author} {\bibfnamefont {D.~A.}\ \bibnamefont
  {Ivanov}},\ }\bibfield  {title} {\enquote {\bibinfo {title} {Non-abelian
  statistics of half-quantum vortices in $\mathit{p}$-wave superconductors},}\
  }\href {\doibase 10.1103/PhysRevLett.86.268} {\bibfield  {journal} {\bibinfo
  {journal} {Phys. Rev. Lett.}\ }\textbf {\bibinfo {volume} {86}},\ \bibinfo
  {pages} {268--271} (\bibinfo {year} {2001})}\BibitemShut {NoStop}%
\bibitem [{\citenamefont {Kitaev}(2003)}]{KITAEV20032}%
  \BibitemOpen
  \bibfield  {author} {\bibinfo {author} {\bibfnamefont {A.Yu.}\ \bibnamefont
  {Kitaev}},\ }\bibfield  {title} {\enquote {\bibinfo {title} {Fault-tolerant
  quantum computation by anyons},}\ }\href {\doibase
  https://doi.org/10.1016/S0003-4916(02)00018-0} {\bibfield  {journal}
  {\bibinfo  {journal} {Annals of Physics}\ }\textbf {\bibinfo {volume}
  {303}},\ \bibinfo {pages} {2--30} (\bibinfo {year} {2003})}\BibitemShut
  {NoStop}%
\bibitem [{\citenamefont {Stern}(2010)}]{Stern2010}%
  \BibitemOpen
  \bibfield  {author} {\bibinfo {author} {\bibfnamefont {Ady}\ \bibnamefont
  {Stern}},\ }\bibfield  {title} {\enquote {\bibinfo {title} {Non-abelian
  states of matter},}\ }\href {\doibase 10.1038/nature08915} {\bibfield
  {journal} {\bibinfo  {journal} {Nature}\ }\textbf {\bibinfo {volume} {464}},\
  \bibinfo {pages} {187--193} (\bibinfo {year} {2010})}\BibitemShut {NoStop}%
\bibitem [{\citenamefont {Nayak}\ \emph {et~al.}(2008)\citenamefont {Nayak},
  \citenamefont {Simon}, \citenamefont {Stern}, \citenamefont {Freedman},\ and\
  \citenamefont {Das~Sarma}}]{CNayak}%
  \BibitemOpen
  \bibfield  {author} {\bibinfo {author} {\bibfnamefont {Chetan}\ \bibnamefont
  {Nayak}}, \bibinfo {author} {\bibfnamefont {Steven~H.}\ \bibnamefont
  {Simon}}, \bibinfo {author} {\bibfnamefont {Ady}\ \bibnamefont {Stern}},
  \bibinfo {author} {\bibfnamefont {Michael}\ \bibnamefont {Freedman}}, \ and\
  \bibinfo {author} {\bibfnamefont {Sankar}\ \bibnamefont {Das~Sarma}},\
  }\bibfield  {title} {\enquote {\bibinfo {title} {Non-abelian anyons and
  topological quantum computation},}\ }\href {\doibase
  10.1103/RevModPhys.80.1083} {\bibfield  {journal} {\bibinfo  {journal} {Rev.
  Mod. Phys.}\ }\textbf {\bibinfo {volume} {80}},\ \bibinfo {pages}
  {1083--1159} (\bibinfo {year} {2008})}\BibitemShut {NoStop}%
\bibitem [{\citenamefont {Kitaev}(2009)}]{Kitaev2009}%
  \BibitemOpen
  \bibfield  {author} {\bibinfo {author} {\bibfnamefont {Alexei}\ \bibnamefont
  {Kitaev}},\ }\bibfield  {title} {\enquote {\bibinfo {title} {Periodic table
  for topological insulators and superconductors},}\ }\href {\doibase
  10.1063/1.3149495} {\bibfield  {journal} {\bibinfo  {journal} {AIP Conference
  Proceedings}\ }\textbf {\bibinfo {volume} {1134}},\ \bibinfo {pages} {22--30}
  (\bibinfo {year} {2009})}\BibitemShut {NoStop}%
\bibitem [{\citenamefont {Aguado}(2017)}]{Aguado2017}%
  \BibitemOpen
  \bibfield  {author} {\bibinfo {author} {\bibfnamefont {Ramón}\ \bibnamefont
  {Aguado}},\ }\bibfield  {title} {\enquote {\bibinfo {title} {Majorana
  quasiparticles in condensed matter},}\ }\href {\doibase
  10.1393/ncr/i2017-10141-9} {\bibfield  {journal} {\bibinfo  {journal} {La
  Rivista del Nuovo Cimento}\ }\textbf {\bibinfo {volume} {40}},\ \bibinfo
  {pages} {523--593} (\bibinfo {year} {2017})}\BibitemShut {NoStop}%
\bibitem [{\citenamefont {Mourik}\ \emph {et~al.}(2012)\citenamefont {Mourik},
  \citenamefont {Zuo}, \citenamefont {Frolov}, \citenamefont {Plissard},
  \citenamefont {Bakkers},\ and\ \citenamefont {Kouwenhoven}}]{Mourik}%
  \BibitemOpen
  \bibfield  {author} {\bibinfo {author} {\bibfnamefont {V.}~\bibnamefont
  {Mourik}}, \bibinfo {author} {\bibfnamefont {K.}~\bibnamefont {Zuo}},
  \bibinfo {author} {\bibfnamefont {S.~M.}\ \bibnamefont {Frolov}}, \bibinfo
  {author} {\bibfnamefont {S.~R.}\ \bibnamefont {Plissard}}, \bibinfo {author}
  {\bibfnamefont {E.~P. A.~M.}\ \bibnamefont {Bakkers}}, \ and\ \bibinfo
  {author} {\bibfnamefont {L.~P.}\ \bibnamefont {Kouwenhoven}},\ }\bibfield
  {title} {\enquote {\bibinfo {title} {Signatures of majorana fermions in
  hybrid superconductor-semiconductor nanowire devices},}\ }\href {\doibase
  10.1126/science.1222360} {\bibfield  {journal} {\bibinfo  {journal}
  {Science}\ }\textbf {\bibinfo {volume} {336}},\ \bibinfo {pages} {1003--1007}
  (\bibinfo {year} {2012})}\BibitemShut {NoStop}%
\bibitem [{\citenamefont {Nadj-Perge}\ \emph {et~al.}(2013)\citenamefont
  {Nadj-Perge}, \citenamefont {Drozdov}, \citenamefont {Bernevig},\ and\
  \citenamefont {Yazdani}}]{Bernevig}%
  \BibitemOpen
  \bibfield  {author} {\bibinfo {author} {\bibfnamefont {S.}~\bibnamefont
  {Nadj-Perge}}, \bibinfo {author} {\bibfnamefont {I.~K.}\ \bibnamefont
  {Drozdov}}, \bibinfo {author} {\bibfnamefont {B.~A.}\ \bibnamefont
  {Bernevig}}, \ and\ \bibinfo {author} {\bibfnamefont {Ali}\ \bibnamefont
  {Yazdani}},\ }\bibfield  {title} {\enquote {\bibinfo {title} {Proposal for
  realizing majorana fermions in chains of magnetic atoms on a
  superconductor},}\ }\href {\doibase 10.1103/PhysRevB.88.020407} {\bibfield
  {journal} {\bibinfo  {journal} {Phys. Rev. B}\ }\textbf {\bibinfo {volume}
  {88}},\ \bibinfo {pages} {020407} (\bibinfo {year} {2013})}\BibitemShut
  {NoStop}%
\bibitem [{\citenamefont {Pientka}\ \emph {et~al.}(2013)\citenamefont
  {Pientka}, \citenamefont {Glazman},\ and\ \citenamefont {von Oppen}}]{Felix}%
  \BibitemOpen
  \bibfield  {author} {\bibinfo {author} {\bibfnamefont {Falko}\ \bibnamefont
  {Pientka}}, \bibinfo {author} {\bibfnamefont {Leonid~I.}\ \bibnamefont
  {Glazman}}, \ and\ \bibinfo {author} {\bibfnamefont {Felix}\ \bibnamefont
  {von Oppen}},\ }\bibfield  {title} {\enquote {\bibinfo {title} {Topological
  superconducting phase in helical shiba chains},}\ }\href {\doibase
  10.1103/PhysRevB.88.155420} {\bibfield  {journal} {\bibinfo  {journal} {Phys.
  Rev. B}\ }\textbf {\bibinfo {volume} {88}},\ \bibinfo {pages} {155420}
  (\bibinfo {year} {2013})}\BibitemShut {NoStop}%
\bibitem [{\citenamefont {Klinovaja}\ \emph {et~al.}(2013)\citenamefont
  {Klinovaja}, \citenamefont {Stano}, \citenamefont {Yazdani},\ and\
  \citenamefont {Loss}}]{LossRKKY}%
  \BibitemOpen
  \bibfield  {author} {\bibinfo {author} {\bibfnamefont {Jelena}\ \bibnamefont
  {Klinovaja}}, \bibinfo {author} {\bibfnamefont {Peter}\ \bibnamefont
  {Stano}}, \bibinfo {author} {\bibfnamefont {Ali}\ \bibnamefont {Yazdani}}, \
  and\ \bibinfo {author} {\bibfnamefont {Daniel}\ \bibnamefont {Loss}},\
  }\bibfield  {title} {\enquote {\bibinfo {title} {Topological
  superconductivity and majorana fermions in rkky systems},}\ }\href {\doibase
  10.1103/PhysRevLett.111.186805} {\bibfield  {journal} {\bibinfo  {journal}
  {Phys. Rev. Lett.}\ }\textbf {\bibinfo {volume} {111}},\ \bibinfo {pages}
  {186805} (\bibinfo {year} {2013})}\BibitemShut {NoStop}%
\bibitem [{\citenamefont {R\"ontynen}\ and\ \citenamefont
  {Ojanen}(2015)}]{Teemu2d}%
  \BibitemOpen
  \bibfield  {author} {\bibinfo {author} {\bibfnamefont {Joel}\ \bibnamefont
  {R\"ontynen}}\ and\ \bibinfo {author} {\bibfnamefont {Teemu}\ \bibnamefont
  {Ojanen}},\ }\bibfield  {title} {\enquote {\bibinfo {title} {Topological
  superconductivity and high chern numbers in 2d ferromagnetic shiba
  lattices},}\ }\href {\doibase 10.1103/PhysRevLett.114.236803} {\bibfield
  {journal} {\bibinfo  {journal} {Phys. Rev. Lett.}\ }\textbf {\bibinfo
  {volume} {114}},\ \bibinfo {pages} {236803} (\bibinfo {year}
  {2015})}\BibitemShut {NoStop}%
\bibitem [{\citenamefont {Hess}\ \emph {et~al.}(2022)\citenamefont {Hess},
  \citenamefont {Legg}, \citenamefont {Loss},\ and\ \citenamefont
  {Klinovaja}}]{LossU1}%
  \BibitemOpen
  \bibfield  {author} {\bibinfo {author} {\bibfnamefont {Richard}\ \bibnamefont
  {Hess}}, \bibinfo {author} {\bibfnamefont {Henry~F.}\ \bibnamefont {Legg}},
  \bibinfo {author} {\bibfnamefont {Daniel}\ \bibnamefont {Loss}}, \ and\
  \bibinfo {author} {\bibfnamefont {Jelena}\ \bibnamefont {Klinovaja}},\
  }\bibfield  {title} {\enquote {\bibinfo {title} {Prevalence of trivial
  zero-energy subgap states in nonuniform helical spin chains on the surface of
  superconductors},}\ }\href {\doibase 10.1103/PhysRevB.106.104503} {\bibfield
  {journal} {\bibinfo  {journal} {Phys. Rev. B}\ }\textbf {\bibinfo {volume}
  {106}},\ \bibinfo {pages} {104503} (\bibinfo {year} {2022})}\BibitemShut
  {NoStop}%
\bibitem [{\citenamefont {Nadj-Perge}\ \emph {et~al.}(2014)\citenamefont
  {Nadj-Perge}, \citenamefont {Drozdov}, \citenamefont {Li}, \citenamefont
  {Chen}, \citenamefont {Jeon}, \citenamefont {Seo}, \citenamefont {MacDonald},
  \citenamefont {Bernevig},\ and\ \citenamefont {Yazdani}}]{SNadjexp}%
  \BibitemOpen
  \bibfield  {author} {\bibinfo {author} {\bibfnamefont {Stevan}\ \bibnamefont
  {Nadj-Perge}}, \bibinfo {author} {\bibfnamefont {Ilya~K.}\ \bibnamefont
  {Drozdov}}, \bibinfo {author} {\bibfnamefont {Jian}\ \bibnamefont {Li}},
  \bibinfo {author} {\bibfnamefont {Hua}\ \bibnamefont {Chen}}, \bibinfo
  {author} {\bibfnamefont {Sangjun}\ \bibnamefont {Jeon}}, \bibinfo {author}
  {\bibfnamefont {Jungpil}\ \bibnamefont {Seo}}, \bibinfo {author}
  {\bibfnamefont {Allan~H.}\ \bibnamefont {MacDonald}}, \bibinfo {author}
  {\bibfnamefont {B.~Andrei}\ \bibnamefont {Bernevig}}, \ and\ \bibinfo
  {author} {\bibfnamefont {Ali}\ \bibnamefont {Yazdani}},\ }\bibfield  {title}
  {\enquote {\bibinfo {title} {Observation of majorana fermions in
  ferromagnetic atomic chains on a superconductor},}\ }\href {\doibase
  10.1126/science.1259327} {\bibfield  {journal} {\bibinfo  {journal}
  {Science}\ }\textbf {\bibinfo {volume} {346}},\ \bibinfo {pages} {602--607}
  (\bibinfo {year} {2014})}\BibitemShut {NoStop}%
\bibitem [{\citenamefont {Kim}\ \emph {et~al.}(2018)\citenamefont {Kim},
  \citenamefont {Palacio-Morales}, \citenamefont {Posske}, \citenamefont
  {Rózsa}, \citenamefont {Palotás}, \citenamefont {Szunyogh}, \citenamefont
  {Thorwart},\ and\ \citenamefont {Wiesendanger}}]{KimSciadv}%
  \BibitemOpen
  \bibfield  {author} {\bibinfo {author} {\bibfnamefont {Howon}\ \bibnamefont
  {Kim}}, \bibinfo {author} {\bibfnamefont {Alexandra}\ \bibnamefont
  {Palacio-Morales}}, \bibinfo {author} {\bibfnamefont {Thore}\ \bibnamefont
  {Posske}}, \bibinfo {author} {\bibfnamefont {Levente}\ \bibnamefont
  {Rózsa}}, \bibinfo {author} {\bibfnamefont {Krisztián}\ \bibnamefont
  {Palotás}}, \bibinfo {author} {\bibfnamefont {László}\ \bibnamefont
  {Szunyogh}}, \bibinfo {author} {\bibfnamefont {Michael}\ \bibnamefont
  {Thorwart}}, \ and\ \bibinfo {author} {\bibfnamefont {Roland}\ \bibnamefont
  {Wiesendanger}},\ }\bibfield  {title} {\enquote {\bibinfo {title} {Toward
  tailoring majorana bound states in artificially constructed magnetic atom
  chains on elemental superconductors},}\ }\href {\doibase
  10.1126/sciadv.aar5251} {\bibfield  {journal} {\bibinfo  {journal} {Science
  Advances}\ }\textbf {\bibinfo {volume} {4}},\ \bibinfo {pages} {eaar5251}
  (\bibinfo {year} {2018})}\BibitemShut {NoStop}%
\bibitem [{\citenamefont {Schneider}\ \emph {et~al.}(2021)\citenamefont
  {Schneider}, \citenamefont {Beck}, \citenamefont {Posske}, \citenamefont
  {Crawford}, \citenamefont {Mascot}, \citenamefont {Rachel}, \citenamefont
  {Wiesendanger},\ and\ \citenamefont {Wiebe}}]{Schneidernat}%
  \BibitemOpen
  \bibfield  {author} {\bibinfo {author} {\bibfnamefont {Lucas}\ \bibnamefont
  {Schneider}}, \bibinfo {author} {\bibfnamefont {Philip}\ \bibnamefont
  {Beck}}, \bibinfo {author} {\bibfnamefont {Thore}\ \bibnamefont {Posske}},
  \bibinfo {author} {\bibfnamefont {Daniel}\ \bibnamefont {Crawford}}, \bibinfo
  {author} {\bibfnamefont {Eric}\ \bibnamefont {Mascot}}, \bibinfo {author}
  {\bibfnamefont {Stephan}\ \bibnamefont {Rachel}}, \bibinfo {author}
  {\bibfnamefont {Roland}\ \bibnamefont {Wiesendanger}}, \ and\ \bibinfo
  {author} {\bibfnamefont {Jens}\ \bibnamefont {Wiebe}},\ }\bibfield  {title}
  {\enquote {\bibinfo {title} {Topological shiba bands in artificial spin
  chains on superconductors},}\ }\href {\doibase 10.1038/s41567-021-01234-y}
  {\bibfield  {journal} {\bibinfo  {journal} {Nature Physics}\ }\textbf
  {\bibinfo {volume} {17}},\ \bibinfo {pages} {943--948} (\bibinfo {year}
  {2021})}\BibitemShut {NoStop}%
\bibitem [{\citenamefont {Crawford}\ \emph {et~al.}(2022)\citenamefont
  {Crawford}, \citenamefont {Mascot}, \citenamefont {Shimizu}, \citenamefont
  {Beck}, \citenamefont {Wiebe}, \citenamefont {Wiesendanger}, \citenamefont
  {Jeschke}, \citenamefont {Morr},\ and\ \citenamefont
  {Rachel}}]{Crawford2022}%
  \BibitemOpen
  \bibfield  {author} {\bibinfo {author} {\bibfnamefont {Daniel}\ \bibnamefont
  {Crawford}}, \bibinfo {author} {\bibfnamefont {Eric}\ \bibnamefont {Mascot}},
  \bibinfo {author} {\bibfnamefont {Makoto}\ \bibnamefont {Shimizu}}, \bibinfo
  {author} {\bibfnamefont {Philip}\ \bibnamefont {Beck}}, \bibinfo {author}
  {\bibfnamefont {Jens}\ \bibnamefont {Wiebe}}, \bibinfo {author}
  {\bibfnamefont {Roland}\ \bibnamefont {Wiesendanger}}, \bibinfo {author}
  {\bibfnamefont {Harald~O.}\ \bibnamefont {Jeschke}}, \bibinfo {author}
  {\bibfnamefont {Dirk~K.}\ \bibnamefont {Morr}}, \ and\ \bibinfo {author}
  {\bibfnamefont {Stephan}\ \bibnamefont {Rachel}},\ }\bibfield  {title}
  {\enquote {\bibinfo {title} {Majorana modes with side features in
  magnet-superconductor hybrid systems},}\ }\href {\doibase
  10.1038/s41535-022-00530-x} {\bibfield  {journal} {\bibinfo  {journal} {npj
  Quantum Materials}\ }\textbf {\bibinfo {volume} {7}},\ \bibinfo {pages} {117}
  (\bibinfo {year} {2022})}\BibitemShut {NoStop}%
\bibitem [{\citenamefont {Schneider}\ \emph {et~al.}(2022)\citenamefont
  {Schneider}, \citenamefont {Beck}, \citenamefont {Neuhaus-Steinmetz},
  \citenamefont {R{\'o}zsa}, \citenamefont {Posske}, \citenamefont {Wiebe},\
  and\ \citenamefont {Wiesendanger}}]{Schneidernat2}%
  \BibitemOpen
  \bibfield  {author} {\bibinfo {author} {\bibfnamefont {Lucas}\ \bibnamefont
  {Schneider}}, \bibinfo {author} {\bibfnamefont {Philip}\ \bibnamefont
  {Beck}}, \bibinfo {author} {\bibfnamefont {Jannis}\ \bibnamefont
  {Neuhaus-Steinmetz}}, \bibinfo {author} {\bibfnamefont {Levente}\
  \bibnamefont {R{\'o}zsa}}, \bibinfo {author} {\bibfnamefont {Thore}\
  \bibnamefont {Posske}}, \bibinfo {author} {\bibfnamefont {Jens}\ \bibnamefont
  {Wiebe}}, \ and\ \bibinfo {author} {\bibfnamefont {Roland}\ \bibnamefont
  {Wiesendanger}},\ }\bibfield  {title} {\enquote {\bibinfo {title} {Precursors
  of majorana modes and their length-dependent energy oscillations probed at
  both ends of atomic shiba chains},}\ }\href {\doibase
  10.1038/s41565-022-01078-4} {\bibfield  {journal} {\bibinfo  {journal}
  {Nature Nanotechnology}\ }\textbf {\bibinfo {volume} {17}},\ \bibinfo {pages}
  {384--389} (\bibinfo {year} {2022})}\BibitemShut {NoStop}%
\bibitem [{\citenamefont {Beck}\ \emph {et~al.}(2021)\citenamefont {Beck},
  \citenamefont {Schneider}, \citenamefont {R{\'o}zsa}, \citenamefont
  {Palot{\'a}s}, \citenamefont {L{\'a}szl{\'o}ffy}, \citenamefont {Szunyogh},
  \citenamefont {Wiebe},\ and\ \citenamefont {Wiesendanger}}]{Beck2021}%
  \BibitemOpen
  \bibfield  {author} {\bibinfo {author} {\bibfnamefont {Philip}\ \bibnamefont
  {Beck}}, \bibinfo {author} {\bibfnamefont {Lucas}\ \bibnamefont {Schneider}},
  \bibinfo {author} {\bibfnamefont {Levente}\ \bibnamefont {R{\'o}zsa}},
  \bibinfo {author} {\bibfnamefont {Kriszti{\'a}n}\ \bibnamefont
  {Palot{\'a}s}}, \bibinfo {author} {\bibfnamefont {Andr{\'a}s}\ \bibnamefont
  {L{\'a}szl{\'o}ffy}}, \bibinfo {author} {\bibfnamefont {L{\'a}szl{\'o}}\
  \bibnamefont {Szunyogh}}, \bibinfo {author} {\bibfnamefont {Jens}\
  \bibnamefont {Wiebe}}, \ and\ \bibinfo {author} {\bibfnamefont {Roland}\
  \bibnamefont {Wiesendanger}},\ }\bibfield  {title} {\enquote {\bibinfo
  {title} {Spin-orbit coupling induced splitting of yu-shiba-rusinov states in
  antiferromagnetic dimers},}\ }\href {\doibase 10.1038/s41467-021-22261-6}
  {\bibfield  {journal} {\bibinfo  {journal} {Nature Communications}\ }\textbf
  {\bibinfo {volume} {12}},\ \bibinfo {pages} {2040} (\bibinfo {year}
  {2021})}\BibitemShut {NoStop}%
\bibitem [{\citenamefont {Fulde}\ and\ \citenamefont
  {Ferrell}(1964)}]{Fulde_1964}%
  \BibitemOpen
  \bibfield  {author} {\bibinfo {author} {\bibfnamefont {Peter}\ \bibnamefont
  {Fulde}}\ and\ \bibinfo {author} {\bibfnamefont {Richard~A.}\ \bibnamefont
  {Ferrell}},\ }\bibfield  {title} {\enquote {\bibinfo {title}
  {Superconductivity in a strong spin-exchange field},}\ }\href {\doibase
  10.1103/PhysRev.135.A550} {\bibfield  {journal} {\bibinfo  {journal} {Phys.
  Rev.}\ }\textbf {\bibinfo {volume} {135}},\ \bibinfo {pages} {A550--A563}
  (\bibinfo {year} {1964})}\BibitemShut {NoStop}%
\bibitem [{\citenamefont {Larkin}\ and\ \citenamefont
  {Ovchinnikov}(1964)}]{Larkin_1964}%
  \BibitemOpen
  \bibfield  {author} {\bibinfo {author} {\bibfnamefont {A.~I.}\ \bibnamefont
  {Larkin}}\ and\ \bibinfo {author} {\bibfnamefont {Y.~N.}\ \bibnamefont
  {Ovchinnikov}},\ }\bibfield  {title} {\enquote {\bibinfo {title} {{Nonuniform
  state of superconductors}},}\ }\href@noop {} {\bibfield  {journal} {\bibinfo
  {journal} {Zh. Eksp. Teor. Fiz.}\ }\textbf {\bibinfo {volume} {47}},\
  \bibinfo {pages} {1136--1146} (\bibinfo {year} {1964})}\BibitemShut {NoStop}%
\bibitem [{\citenamefont {Yuan}\ and\ \citenamefont {Fu}(2022)}]{LiangPNAS}%
  \BibitemOpen
  \bibfield  {author} {\bibinfo {author} {\bibfnamefont {Noah F.~Q.}\
  \bibnamefont {Yuan}}\ and\ \bibinfo {author} {\bibfnamefont {Liang}\
  \bibnamefont {Fu}},\ }\bibfield  {title} {\enquote {\bibinfo {title}
  {Supercurrent diode effect and finite-momentum superconductors},}\ }\href
  {\doibase 10.1073/pnas.2119548119} {\bibfield  {journal} {\bibinfo  {journal}
  {Proceedings of the National Academy of Sciences}\ }\textbf {\bibinfo
  {volume} {119}},\ \bibinfo {pages} {e2119548119} (\bibinfo {year}
  {2022})}\BibitemShut {NoStop}%
\bibitem [{\citenamefont {Ili\ifmmode~\acute{c}\else \'{c}\fi{}}\ and\
  \citenamefont {Bergeret}(2022)}]{SLlicprl}%
  \BibitemOpen
  \bibfield  {author} {\bibinfo {author} {\bibfnamefont {S.}~\bibnamefont
  {Ili\ifmmode~\acute{c}\else \'{c}\fi{}}}\ and\ \bibinfo {author}
  {\bibfnamefont {F.~S.}\ \bibnamefont {Bergeret}},\ }\bibfield  {title}
  {\enquote {\bibinfo {title} {Theory of the supercurrent diode effect in
  rashba superconductors with arbitrary disorder},}\ }\href {\doibase
  10.1103/PhysRevLett.128.177001} {\bibfield  {journal} {\bibinfo  {journal}
  {Phys. Rev. Lett.}\ }\textbf {\bibinfo {volume} {128}},\ \bibinfo {pages}
  {177001} (\bibinfo {year} {2022})}\BibitemShut {NoStop}%
\bibitem [{\citenamefont {Daido}\ and\ \citenamefont
  {Yanase}(2022)}]{Yanasediode}%
  \BibitemOpen
  \bibfield  {author} {\bibinfo {author} {\bibfnamefont {Akito}\ \bibnamefont
  {Daido}}\ and\ \bibinfo {author} {\bibfnamefont {Youichi}\ \bibnamefont
  {Yanase}},\ }\bibfield  {title} {\enquote {\bibinfo {title} {Superconducting
  diode effect and nonreciprocal transition lines},}\ }\href {\doibase
  10.1103/PhysRevB.106.205206} {\bibfield  {journal} {\bibinfo  {journal}
  {Phys. Rev. B}\ }\textbf {\bibinfo {volume} {106}},\ \bibinfo {pages}
  {205206} (\bibinfo {year} {2022})}\BibitemShut {NoStop}%
\bibitem [{\citenamefont {de~Picoli}\ \emph {et~al.}(2023)\citenamefont
  {de~Picoli}, \citenamefont {Blood}, \citenamefont {Lyanda-Geller},\ and\
  \citenamefont {V\"ayrynen}}]{Picoli}%
  \BibitemOpen
  \bibfield  {author} {\bibinfo {author} {\bibfnamefont {Tatiana}\ \bibnamefont
  {de~Picoli}}, \bibinfo {author} {\bibfnamefont {Zane}\ \bibnamefont {Blood}},
  \bibinfo {author} {\bibfnamefont {Yuli}\ \bibnamefont {Lyanda-Geller}}, \
  and\ \bibinfo {author} {\bibfnamefont {Jukka~I.}\ \bibnamefont
  {V\"ayrynen}},\ }\bibfield  {title} {\enquote {\bibinfo {title}
  {Superconducting diode effect in quasi-one-dimensional systems},}\ }\href
  {\doibase 10.1103/PhysRevB.107.224518} {\bibfield  {journal} {\bibinfo
  {journal} {Phys. Rev. B}\ }\textbf {\bibinfo {volume} {107}},\ \bibinfo
  {pages} {224518} (\bibinfo {year} {2023})}\BibitemShut {NoStop}%
\bibitem [{\citenamefont {He}\ \emph {et~al.}(2022)\citenamefont {He},
  \citenamefont {Tanaka},\ and\ \citenamefont {Nagaosa}}]{Nagaosanjp}%
  \BibitemOpen
  \bibfield  {author} {\bibinfo {author} {\bibfnamefont {James~Jun}\
  \bibnamefont {He}}, \bibinfo {author} {\bibfnamefont {Yukio}\ \bibnamefont
  {Tanaka}}, \ and\ \bibinfo {author} {\bibfnamefont {Naoto}\ \bibnamefont
  {Nagaosa}},\ }\bibfield  {title} {\enquote {\bibinfo {title} {A
  phenomenological theory of superconductor diodes},}\ }\href {\doibase
  10.1088/1367-2630/ac6766} {\bibfield  {journal} {\bibinfo  {journal} {New
  Journal of Physics}\ }\textbf {\bibinfo {volume} {24}},\ \bibinfo {pages}
  {053014} (\bibinfo {year} {2022})}\BibitemShut {NoStop}%
\bibitem [{\citenamefont {Qu}\ \emph {et~al.}(2013)\citenamefont {Qu},
  \citenamefont {Zheng}, \citenamefont {Gong}, \citenamefont {Xu},
  \citenamefont {Mao}, \citenamefont {Zou}, \citenamefont {Guo},\ and\
  \citenamefont {Zhang}}]{Qu2013natcom}%
  \BibitemOpen
  \bibfield  {author} {\bibinfo {author} {\bibfnamefont {Chunlei}\ \bibnamefont
  {Qu}}, \bibinfo {author} {\bibfnamefont {Zhen}\ \bibnamefont {Zheng}},
  \bibinfo {author} {\bibfnamefont {Ming}\ \bibnamefont {Gong}}, \bibinfo
  {author} {\bibfnamefont {Yong}\ \bibnamefont {Xu}}, \bibinfo {author}
  {\bibfnamefont {Li}~\bibnamefont {Mao}}, \bibinfo {author} {\bibfnamefont
  {Xubo}\ \bibnamefont {Zou}}, \bibinfo {author} {\bibfnamefont {Guangcan}\
  \bibnamefont {Guo}}, \ and\ \bibinfo {author} {\bibfnamefont {Chuanwei}\
  \bibnamefont {Zhang}},\ }\bibfield  {title} {\enquote {\bibinfo {title}
  {Topological superfluids with finite-momentum pairing and majorana
  fermions},}\ }\href {\doibase 10.1038/ncomms3710} {\bibfield  {journal}
  {\bibinfo  {journal} {Nature Communications}\ }\textbf {\bibinfo {volume}
  {4}},\ \bibinfo {pages} {2710} (\bibinfo {year} {2013})}\BibitemShut
  {NoStop}%
\bibitem [{\citenamefont {Nadeem}\ \emph {et~al.}(2023)\citenamefont {Nadeem},
  \citenamefont {Fuhrer},\ and\ \citenamefont {Wang}}]{Nadeem2023}%
  \BibitemOpen
  \bibfield  {author} {\bibinfo {author} {\bibfnamefont {Muhammad}\
  \bibnamefont {Nadeem}}, \bibinfo {author} {\bibfnamefont {Michael~S}\
  \bibnamefont {Fuhrer}}, \ and\ \bibinfo {author} {\bibfnamefont {Xiaolin}\
  \bibnamefont {Wang}},\ }\bibfield  {title} {\enquote {\bibinfo {title} {The
  superconducting diode effect},}\ }\href
  {https://doi.org/10.1038/s42254-023-00632-w} {\bibfield  {journal} {\bibinfo
  {journal} {Nature Reviews Physics}\ }\textbf {\bibinfo {volume} {5}},\
  \bibinfo {pages} {558--577} (\bibinfo {year} {2023})}\BibitemShut {NoStop}%
\bibitem [{\citenamefont {Legg}\ \emph {et~al.}(2022)\citenamefont {Legg},
  \citenamefont {Loss},\ and\ \citenamefont {Klinovaja}}]{Lossdiode}%
  \BibitemOpen
  \bibfield  {author} {\bibinfo {author} {\bibfnamefont {Henry~F.}\
  \bibnamefont {Legg}}, \bibinfo {author} {\bibfnamefont {Daniel}\ \bibnamefont
  {Loss}}, \ and\ \bibinfo {author} {\bibfnamefont {Jelena}\ \bibnamefont
  {Klinovaja}},\ }\bibfield  {title} {\enquote {\bibinfo {title}
  {Superconducting diode effect due to magnetochiral anisotropy in topological
  insulators and rashba nanowires},}\ }\href {\doibase
  10.1103/PhysRevB.106.104501} {\bibfield  {journal} {\bibinfo  {journal}
  {Phys. Rev. B}\ }\textbf {\bibinfo {volume} {106}},\ \bibinfo {pages}
  {104501} (\bibinfo {year} {2022})}\BibitemShut {NoStop}%
\bibitem [{\citenamefont {Daido}\ \emph {et~al.}(2022)\citenamefont {Daido},
  \citenamefont {Ikeda},\ and\ \citenamefont {Yanase}}]{Yanaseprl}%
  \BibitemOpen
  \bibfield  {author} {\bibinfo {author} {\bibfnamefont {Akito}\ \bibnamefont
  {Daido}}, \bibinfo {author} {\bibfnamefont {Yuhei}\ \bibnamefont {Ikeda}}, \
  and\ \bibinfo {author} {\bibfnamefont {Youichi}\ \bibnamefont {Yanase}},\
  }\bibfield  {title} {\enquote {\bibinfo {title} {Intrinsic superconducting
  diode effect},}\ }\href {\doibase 10.1103/PhysRevLett.128.037001} {\bibfield
  {journal} {\bibinfo  {journal} {Phys. Rev. Lett.}\ }\textbf {\bibinfo
  {volume} {128}},\ \bibinfo {pages} {037001} (\bibinfo {year}
  {2022})}\BibitemShut {NoStop}%
\bibitem [{\citenamefont {Bhowmik}\ \emph {et~al.}()\citenamefont {Bhowmik},
  \citenamefont {Samanta}, \citenamefont {Nandy}, \citenamefont {Saha},\ and\
  \citenamefont {Ghosh}}]{sayakarxiv}%
  \BibitemOpen
  \bibfield  {author} {\bibinfo {author} {\bibfnamefont {Sayak}\ \bibnamefont
  {Bhowmik}}, \bibinfo {author} {\bibfnamefont {Dibyendu}\ \bibnamefont
  {Samanta}}, \bibinfo {author} {\bibfnamefont {Ashis~K.}\ \bibnamefont
  {Nandy}}, \bibinfo {author} {\bibfnamefont {Arijit}\ \bibnamefont {Saha}}, \
  and\ \bibinfo {author} {\bibfnamefont {Sudeep~Kumar}\ \bibnamefont {Ghosh}},\
  }\href@noop {} {\enquote {\bibinfo {title} {Optimizing one dimensional
  superconducting diodes: Interplay of rashba spin-orbit coupling and magnetic
  fields},}\ }\Eprint {http://arxiv.org/abs/2407.12455} {arXiv:2407.12455
  [cond-mat.supr-con]} \BibitemShut {NoStop}%
\bibitem [{\citenamefont {Nandy}\ \emph {et~al.}(2016)\citenamefont {Nandy},
  \citenamefont {Kiselev},\ and\ \citenamefont {Bl\"ugel}}]{SStexture_ANandy}%
  \BibitemOpen
  \bibfield  {author} {\bibinfo {author} {\bibfnamefont {Ashis~Kumar}\
  \bibnamefont {Nandy}}, \bibinfo {author} {\bibfnamefont {Nikolai~S.}\
  \bibnamefont {Kiselev}}, \ and\ \bibinfo {author} {\bibfnamefont {Stefan}\
  \bibnamefont {Bl\"ugel}},\ }\bibfield  {title} {\enquote {\bibinfo {title}
  {Interlayer exchange coupling: A general scheme turning chiral magnets into
  magnetic multilayers carrying atomic-scale skyrmions},}\ }\href {\doibase
  10.1103/PhysRevLett.116.177202} {\bibfield  {journal} {\bibinfo  {journal}
  {Phys. Rev. Lett.}\ }\textbf {\bibinfo {volume} {116}},\ \bibinfo {pages}
  {177202} (\bibinfo {year} {2016})}\BibitemShut {NoStop}%
\bibitem [{\citenamefont {Chatterjee}\ \emph {et~al.}(2023)\citenamefont
  {Chatterjee}, \citenamefont {Pradhan}, \citenamefont {Nandy},\ and\
  \citenamefont {Saha}}]{Pritamprb1}%
  \BibitemOpen
  \bibfield  {author} {\bibinfo {author} {\bibfnamefont {Pritam}\ \bibnamefont
  {Chatterjee}}, \bibinfo {author} {\bibfnamefont {Saurabh}\ \bibnamefont
  {Pradhan}}, \bibinfo {author} {\bibfnamefont {Ashis~K.}\ \bibnamefont
  {Nandy}}, \ and\ \bibinfo {author} {\bibfnamefont {Arijit}\ \bibnamefont
  {Saha}},\ }\bibfield  {title} {\enquote {\bibinfo {title} {Tailoring the
  phase transition from topological superconductor to trivial superconductor
  induced by magnetic textures of a spin chain on a $p$-wave superconductor},}\
  }\href {\doibase 10.1103/PhysRevB.107.085423} {\bibfield  {journal} {\bibinfo
   {journal} {Phys. Rev. B}\ }\textbf {\bibinfo {volume} {107}},\ \bibinfo
  {pages} {085423} (\bibinfo {year} {2023})}\BibitemShut {NoStop}%
\bibitem [{\citenamefont {Chatterjee}\ \emph
  {et~al.}(2024{\natexlab{a}})\citenamefont {Chatterjee}, \citenamefont
  {Banik}, \citenamefont {Bera}, \citenamefont {Ghosh}, \citenamefont
  {Pradhan}, \citenamefont {Saha},\ and\ \citenamefont {Nandy}}]{Pritamprbl}%
  \BibitemOpen
  \bibfield  {author} {\bibinfo {author} {\bibfnamefont {Pritam}\ \bibnamefont
  {Chatterjee}}, \bibinfo {author} {\bibfnamefont {Sayan}\ \bibnamefont
  {Banik}}, \bibinfo {author} {\bibfnamefont {Sandip}\ \bibnamefont {Bera}},
  \bibinfo {author} {\bibfnamefont {Arnob~Kumar}\ \bibnamefont {Ghosh}},
  \bibinfo {author} {\bibfnamefont {Saurabh}\ \bibnamefont {Pradhan}}, \bibinfo
  {author} {\bibfnamefont {Arijit}\ \bibnamefont {Saha}}, \ and\ \bibinfo
  {author} {\bibfnamefont {Ashis~K.}\ \bibnamefont {Nandy}},\ }\bibfield
  {title} {\enquote {\bibinfo {title} {Topological superconductivity by
  engineering noncollinear magnetism in magnet/superconductor heterostructures:
  A realistic prescription for the two-dimensional kitaev model},}\ }\href
  {\doibase 10.1103/PhysRevB.109.L121301} {\bibfield  {journal} {\bibinfo
  {journal} {Phys. Rev. B}\ }\textbf {\bibinfo {volume} {109}},\ \bibinfo
  {pages} {L121301} (\bibinfo {year} {2024}{\natexlab{a}})}\BibitemShut
  {NoStop}%
\bibitem [{\citenamefont {Chatterjee}\ \emph
  {et~al.}(2024{\natexlab{b}})\citenamefont {Chatterjee}, \citenamefont
  {Ghosh}, \citenamefont {Nandy},\ and\ \citenamefont {Saha}}]{Pritamprb}%
  \BibitemOpen
  \bibfield  {author} {\bibinfo {author} {\bibfnamefont {Pritam}\ \bibnamefont
  {Chatterjee}}, \bibinfo {author} {\bibfnamefont {Arnob~Kumar}\ \bibnamefont
  {Ghosh}}, \bibinfo {author} {\bibfnamefont {Ashis~K.}\ \bibnamefont {Nandy}},
  \ and\ \bibinfo {author} {\bibfnamefont {Arijit}\ \bibnamefont {Saha}},\
  }\bibfield  {title} {\enquote {\bibinfo {title} {Second-order topological
  superconductor via noncollinear magnetic texture},}\ }\href {\doibase
  10.1103/PhysRevB.109.L041409} {\bibfield  {journal} {\bibinfo  {journal}
  {Phys. Rev. B}\ }\textbf {\bibinfo {volume} {109}},\ \bibinfo {pages}
  {L041409} (\bibinfo {year} {2024}{\natexlab{b}})}\BibitemShut {NoStop}%
\bibitem [{sup()}]{supp}%
  \BibitemOpen
  \href@noop {} {}\bibinfo {note} {See the Supplementary Material at
  XXXXXXXXXXX for the details of self-consistent analysis in real space and
  topological superconductivity considering uniform superconducting
  gap.}\BibitemShut {Stop}%
\bibitem [{\citenamefont {Resta}(1998)}]{Resta}%
  \BibitemOpen
  \bibfield  {author} {\bibinfo {author} {\bibfnamefont {Raffaele}\
  \bibnamefont {Resta}},\ }\bibfield  {title} {\enquote {\bibinfo {title}
  {Quantum-mechanical position operator in extended systems},}\ }\href
  {\doibase 10.1103/PhysRevLett.80.1800} {\bibfield  {journal} {\bibinfo
  {journal} {Phys. Rev. Lett.}\ }\textbf {\bibinfo {volume} {80}},\ \bibinfo
  {pages} {1800--1803} (\bibinfo {year} {1998})}\BibitemShut {NoStop}%
\bibitem [{\citenamefont {Wheeler}\ \emph {et~al.}(2019)\citenamefont
  {Wheeler}, \citenamefont {Wagner},\ and\ \citenamefont {Hughes}}]{Wheeler}%
  \BibitemOpen
  \bibfield  {author} {\bibinfo {author} {\bibfnamefont {William~A.}\
  \bibnamefont {Wheeler}}, \bibinfo {author} {\bibfnamefont {Lucas~K.}\
  \bibnamefont {Wagner}}, \ and\ \bibinfo {author} {\bibfnamefont {Taylor~L.}\
  \bibnamefont {Hughes}},\ }\bibfield  {title} {\enquote {\bibinfo {title}
  {Many-body electric multipole operators in extended systems},}\ }\href
  {\doibase 10.1103/PhysRevB.100.245135} {\bibfield  {journal} {\bibinfo
  {journal} {Phys. Rev. B}\ }\textbf {\bibinfo {volume} {100}},\ \bibinfo
  {pages} {245135} (\bibinfo {year} {2019})}\BibitemShut {NoStop}%
\bibitem [{\citenamefont {Kang}\ \emph {et~al.}(2019)\citenamefont {Kang},
  \citenamefont {Shiozaki},\ and\ \citenamefont {Cho}}]{KangPRB}%
  \BibitemOpen
  \bibfield  {author} {\bibinfo {author} {\bibfnamefont {Byungmin}\
  \bibnamefont {Kang}}, \bibinfo {author} {\bibfnamefont {Ken}\ \bibnamefont
  {Shiozaki}}, \ and\ \bibinfo {author} {\bibfnamefont {Gil~Young}\
  \bibnamefont {Cho}},\ }\bibfield  {title} {\enquote {\bibinfo {title}
  {Many-body order parameters for multipoles in solids},}\ }\href {\doibase
  10.1103/PhysRevB.100.245134} {\bibfield  {journal} {\bibinfo  {journal}
  {Phys. Rev. B}\ }\textbf {\bibinfo {volume} {100}},\ \bibinfo {pages}
  {245134} (\bibinfo {year} {2019})}\BibitemShut {NoStop}%
\end{thebibliography}%

\clearpage

\newpage

\begin{onecolumngrid}
\begin{center}

{
	\fontsize{12}{12}
	\selectfont
	\textbf{Supplemental material for ``Topological Majorana zero modes and the superconducting diode effect driven by Fulde-Ferrell-Larkin-Ovchinnikov pairing in a helical Shiba chain''
	\\[5mm]}
}
\normalsize Sayak Bhowmik$^{1,2}$, and Arijit Saha$^{1,2}$\\
\vspace{2mm}
{\small $^1$\textit{Institute of Physics, Sachivalaya Marg, Bhubaneswar-751005, India}\\[0.5mm]}
{\small $^2$\textit{Homi Bhabha National Institute, Training School Complex, Anushakti Nagar, Mumbai 400094, India}\\[0.5mm]}

\end{center}

\renewcommand{\thesection}{S\arabic{section}}

\tableofcontents 

\section{Appearance of FFLO pairing considering lattice model (details of self-consistent analysis) \label{sec:sec_1}}
In this section we demonstrate the self-consistent solution of $s$-wave Fulde-Ferrell-Larkin-Ovchinnikov (FFLO) order parameter: $\Delta(q)$ and the ground state Cooper pair momentum: $q_0$ 
by considering the real space model represented by the one-dimensional (1D) lattice Hamiltonian with open boundary condition (OBC). The mean-field Hamiltonian is given as
\begin{eqnarray}
&&\mathcal{H}= H_l+  \frac{L}{U} |\Delta|^2, \nonumber\\  
&& H_l= \! - t \!\sum_{\langle m,n \rangle,s}   c_{m,s}^{\dagger}  c_{n,s} \! - \sum_{n} \Delta e^{iqn}
(c_{n,\uparrow}^{\dagger}c_{n,\downarrow}^{\dagger}\!+\! {\rm H.c.})  +J  \! \sum_{n,s,{s}^\prime} \! \! c_{n,s}^{\dagger} \! \left(\vect{S}_n \! \cdot \! \vect{\sigma}\right)_{s, {s}^\prime} c_{n,{s}^\prime}   \nonumber\\  
&& \quad\quad-\mu \sum_{n,s} c_{n,s}^{\dagger} c_{n,s} \! \! + B_z  \! \sum_{n,s,{s}^\prime} \! \! c_{n,s}^{\dagger} (\sigma_z)_{s, {s}^\prime} c_{n,{s}^\prime}\ . \quad ~
\label{Slattice}\ 
\end{eqnarray}
where, $t$ represents the hopping amplitude between nearest neighbor sites $\langle m,n \rangle$, and $S_n = \begin{pmatrix} \cos[\phi(n)], & \sin[\phi(n)], & 0 \end{pmatrix}$ denotes the local 
classical spin vector at site $n$, providing the realization of the spin texture in the lattice, where $\phi(n+1) - \phi(n) = g$. The remaining model parameters are defined as mentioned in Eq.~(1) 
of the main text. The FFLO superconducting (SC) order parameter $\Delta(q)$ is self-consistently determined by minimizing the condensation energy: $\Omega(q,\Delta) = F(q,\Delta) - F(q,0)$ 
for a particular value of $q$. Here, $F(q,\Delta)$ denotes the free energy density corresponding to the Hamiltonian in Eq.~(\ref{Slattice}), given as
\beq
F(q,\Delta) = -\frac{1}{L\beta}\sum_{ni} \ln\left[1+e^{-\beta E_{ni}(q)}\right] + \frac{|\Delta(q)|^2}{U}\ ,
\label{eqn:S3}
\eeq
\begin{figure}[t!]
\centering \includegraphics[width=0.85\linewidth] {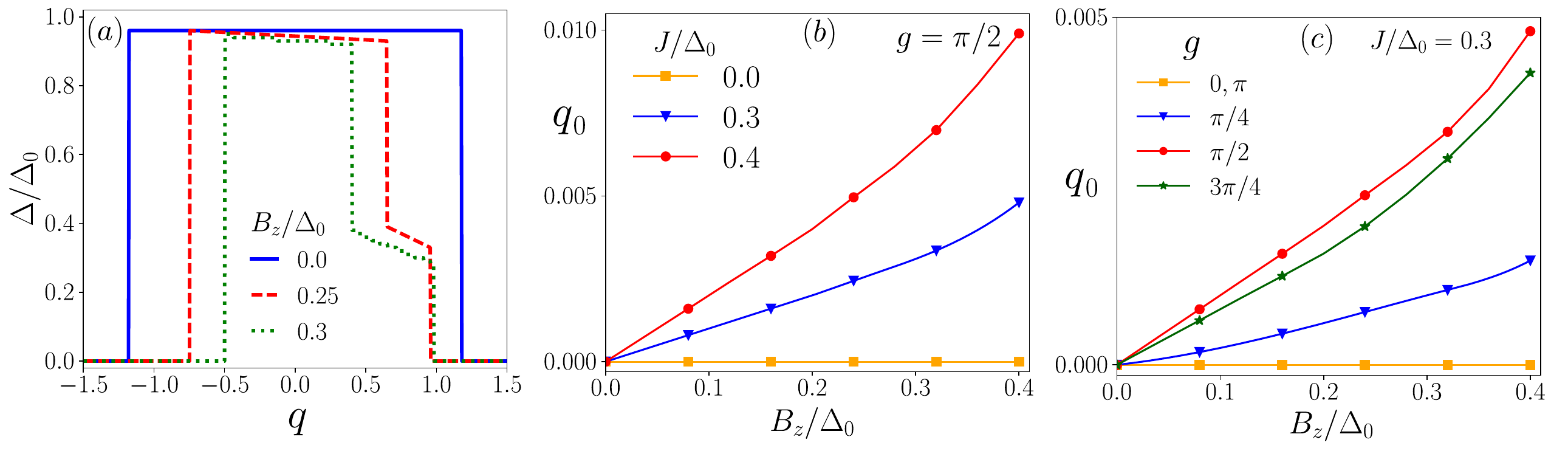}
\caption{\textbf{FFLO pairing stability and ground state:} Panel (a) exhibits the self-consistent superconducting gap $\Delta(q)$ as a function of $q$ choosing different values of external field $B_z$  
with  $J/\Delta_0=0.65$ fixed. The $\Delta(q)$ profile is symmetric with respect to $q$ when $B_z=0$. However, presence of non zero $B_z$ results in asymmetry of $\Delta(q)$ profile. Panel (b) demonstrates the FFLO ground state Cooper pair momentum $q_0$ as a function of the external field $B_z$ for various values of $J$ considering $g=\pi/2$. Panel (c) illustrates $q_0$ as a function of $B_z$ for various values of $g$ (representing different configurations of the spin texture). The results are obtained for a 1D lattice with 400 sites while the other lattice model parameters are chosen 
as $(t/\Delta_0, \mu/\Delta_0, \beta^{-1}, U)=(0.5,1.0, 0.01~\text{meV},2.65~\text{meV})$. }\label{fig:FigS1}
\end{figure}
where, $\beta=(k_B T)^{-1}$ with  $T$ being the temperature and $k_B$ the Boltzmann constant and $ni$ denotes the Bogoliubov-de-Gennes (BdG) quasiparticle eigenvalue index. Having obtained 
$\Delta(q)$  the $s$-wave FFLO stability [see Fig.~\ref{fig:FigS1}(a)], we further determine the FFLO superconducting ground state Cooper pair momentum: $q_0$ by optimizing $\Omega(q,\Delta)$ 
with respect to $q$ by solving: 
\beq
\frac{\delta \Omega(q,\Delta) }{\delta q}\bigg|_{q=q_0}=0\ .
\label{eqn:Sground}
\eeq   
As shown in Fig.~\ref{fig:FigS1}(b), a finite $q_0$ is generated in the 1D Shiba chain featuring spin spiral (SS) (here out-of-plane N\'eel-type) when the external field $B_z$ is non-zero. For trivial configuration of the spin texture ($g=0,\pi$) representing a ferromagnetic, anti-ferromagnetic configurations respectively, $q_0$ remains zero even for a finite $B_z$ [see Fig.~\ref{fig:FigS1}(c)]. As a result, the choice of this trivial configurations of the SS doesn't contribute to the superconducting diode effect (SDE). The above results, obtained based on consideration of the lattice Hamiltonian [Eq.~(\ref{Slattice})] and depicted in Fig.~\ref{fig:FigS1}, are in strong qualitative agreement with those obtained from the continuum model Hamiltonian as discussed in the main text
[see Fig.~2 in the main text].


\section{Topological superconducting phase considering uniform superconducting gap $\Delta_c$ \label{sec:sec_2}}

In this section, we present the results using a constant superconducting gap $\Delta_c$ across the entire system parameter regime, without enforcing self-consistency. The variation of the $\Delta$ 
profile with respect to continuously changing system parameters is not considered; instead, a uniform gap is assumed. 
\begin{figure}[b!]
\centering \includegraphics[width=0.7\linewidth] {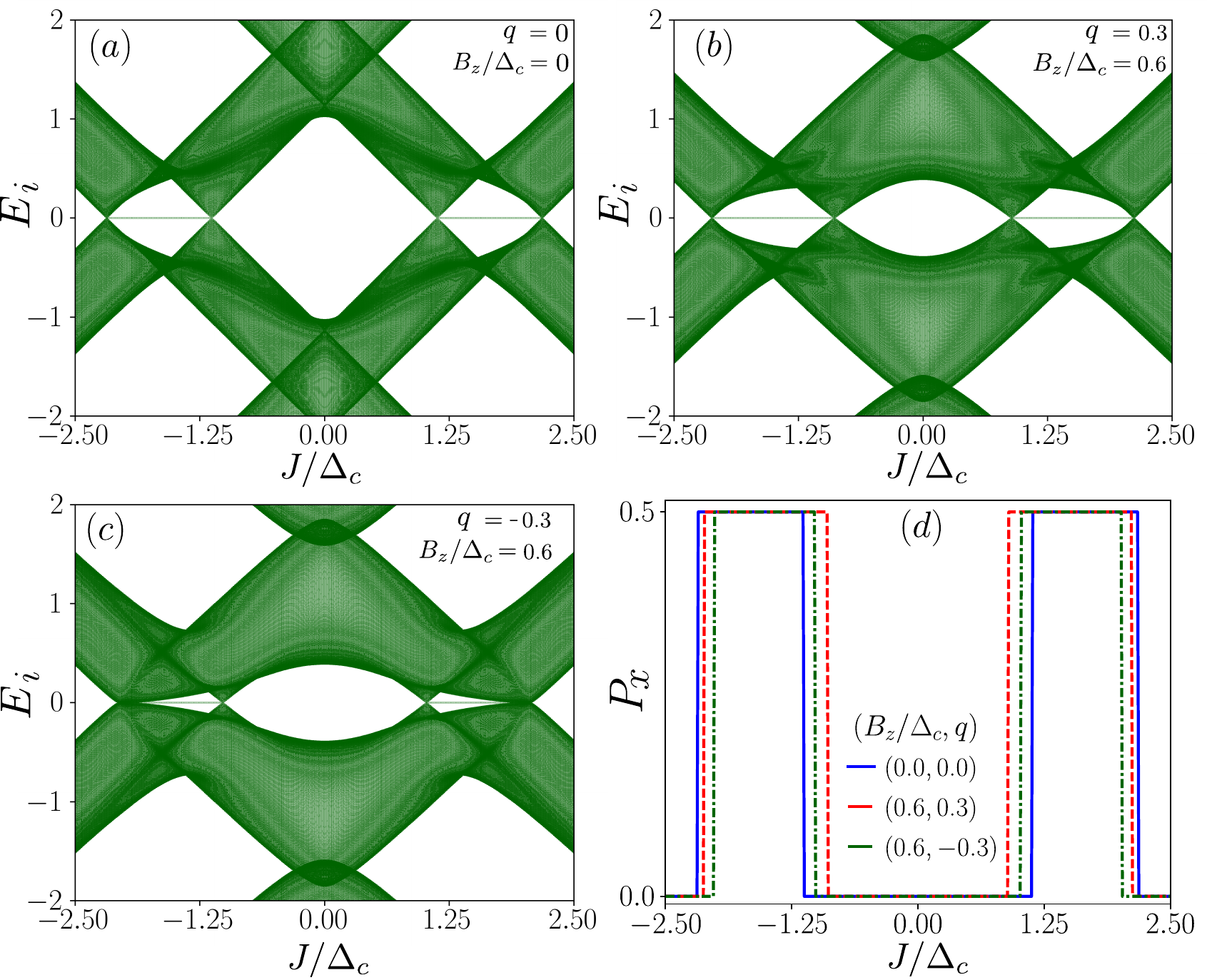}
\caption{\textbf{Gap closing and re-opening transitions}: The OBC energy eigenvalue spectrum $E_i$ of the 1D lattice is visually depicted as a function of the coupling constant $J$  in panel (a) 
for $B_z/\Delta_c=0.0, q=0$ (absence of FFLO pairing), in panel (b) for $B_z/\Delta_c=0.6, q=0.3 $ and in panel (c) for $B_z/\Delta_c=0.6, q=-0.3$. The corresponding topological invariant $P_x$ 
is shown as a function of $J$ in panel (d). Other lattice parameters are chosen as $(t/\Delta_c, \mu/\Delta_c,g, \Delta_c)=(0.5,1.0,\pi/2, 1.0~\text{meV} )$. We choose a finite system size of 
400 lattice sites.}
\label{fig:FigS2}
\end{figure}
This simplification is commonly used in studies exploring the emergence of Majorana zero modes (MZMs) in topological superconductors~\cite{Bernevig, Pritamprbl,Pritamprb}.  However, it leads to qualitative differences in the topological phase diagrams and the behavior of gap-closing transitions when compared to results derived from a self-consistent treatment of the superconducting gap. The topological signatures derived by incorporating the self-consistent solution for $\Delta$ across different parameter sets have already been discussed in the main text (see Fig.~3 of the main text). As shown in Fig.~\ref{fig:FigS2}(a)-(c), the Bogoliubov-de-Genes (BdG) quasiparticle spectrum udergoes gap-closing and re-opening transitions when investigated as a function of $J$ (exchange coupling strength between the spin superconducting electrons and the magnetic impurity atoms) and uniform $\Delta_c$. The presence of FFLO pairing when $B_z\ne0$ (leading to $q\ne0$), only modifies the topological regime. They are characterized by the appropriate topological invariant (polarization $P_x$) as shown in Fig.~\ref{fig:FigS2}(d). 

\begin{figure}[t!]
\centering \includegraphics[width=0.75\linewidth] {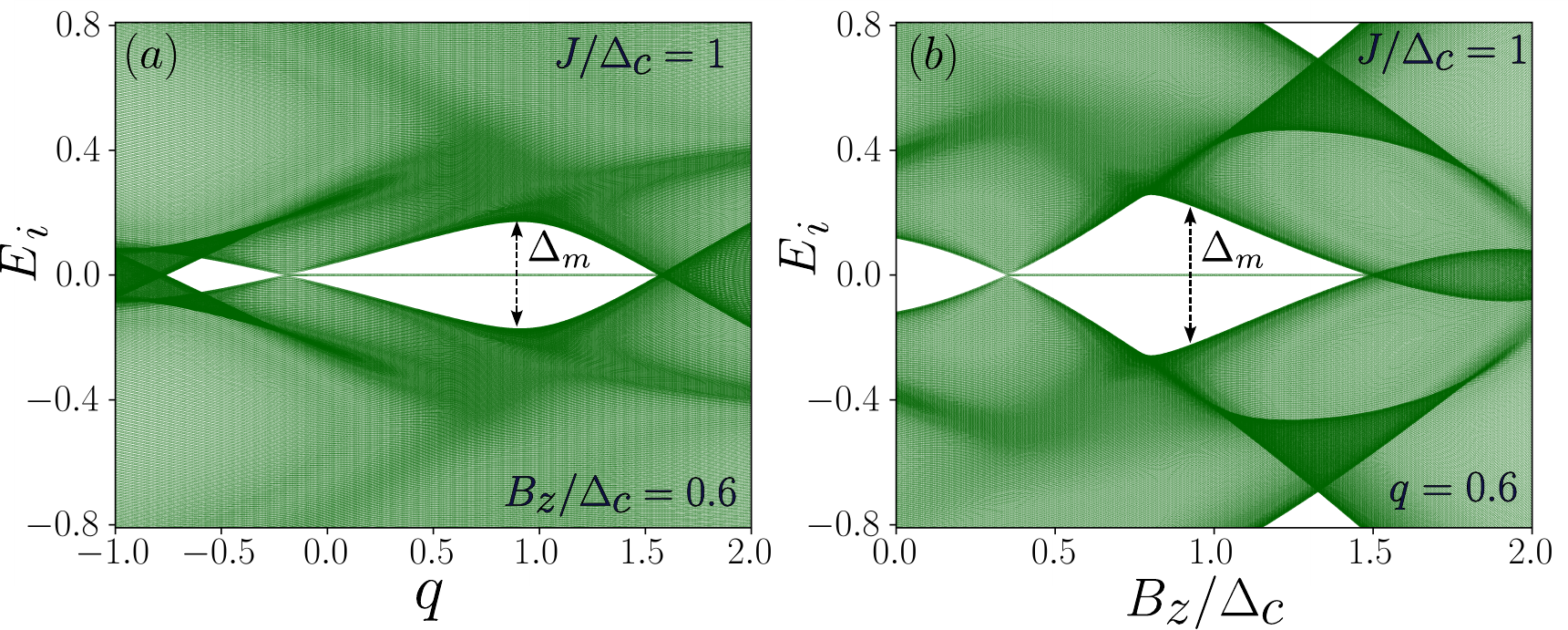}
\caption{\textbf{FFLO pairing induced gapped - gapless phase transitions}: (a) The energy eigenvalue spectrum $E_i$ of the lattice considering 400 sites is shown as a function of the Cooper pair momentum $q$ choosing $B_z/\Delta_c=0.6, J/\Delta_c=1$. (b) $E_i$ is displayed as a function of $B_z$ for $q=0.6, J/\Delta_c=1$. Here $\Delta_m$ denotes the effective mini-gap that separates the MZMs from the bulk.  Other model parameters are chosen as $(t/\Delta_c, \mu/\Delta_c,g, \Delta_c)=(0.5,1.0, \pi/2,1.0~\text{meV})$.}
\label{fig:FigS3}
\end{figure}

The BdG quasiparticle eigenvalue spectrum exhibits smooth gap closing transitions as a fucntion of $q$, with MZM signatures protected by the effective minigap $\Delta_m$, ultimately leading to a 
gap-less phase [see Figs.~\ref{fig:FigS3}(a)-(b)]. Furthermore, the bulk BdG spectrum is asymmetric with respect to $q=0$, as expected, due to the nonreciprocal nature of the system in the presence 
of FFLO pairing. This is evident from Fig.~\ref{fig:FigS3}(a). The bulk BdG spectrum is shown as a function of $B_{z}$ in Fig.~\ref{fig:FigS3}(b) choosing $q\neq 0$ and similar minigap $\Delta_m$ protected MZMs appear in the topological superconduting phase. 

The effective minigap $\Delta_m$ is portrayed in $B_z-q$ plane choosing different values of $g$ [see Figs.~\ref{fig:FigS4}(a),(c)]. The corresponding bulk-dipole moment $P_x$ (topological invariant) profile is depicted in Figs.~\ref{fig:FigS4}(b),(d) considering the same parameter plane. The presence of a finite minigap correctly identifies the topological regime with the presence of MZMs as consistent with the bulk-dipole moment $P_x$ profile.
Furthermore, it is evident from the minigap $\Delta_m$ and $P_x$ profile in $B_z-q$ plane [see Fig.~\ref{fig:FigS4}], that the Cooper pair momentum $q$ in the FFLO state intriguingly assists transitions from topologically trivial to a topological superconducting state. Therefore, incorporating self-consistency in the solution of $\Delta$ across different parameter sets exhibits similar physics, however 
the corresponding phase diagram for $\Delta_m$ and $P_x$ in $B_z-q$ plane appear to be quantitatively different (see Fig.~3 of the main text) than that of the uniform superconducting gap $\Delta_c$.
\begin{figure}[t!]
\centering \includegraphics[width=0.72\linewidth] {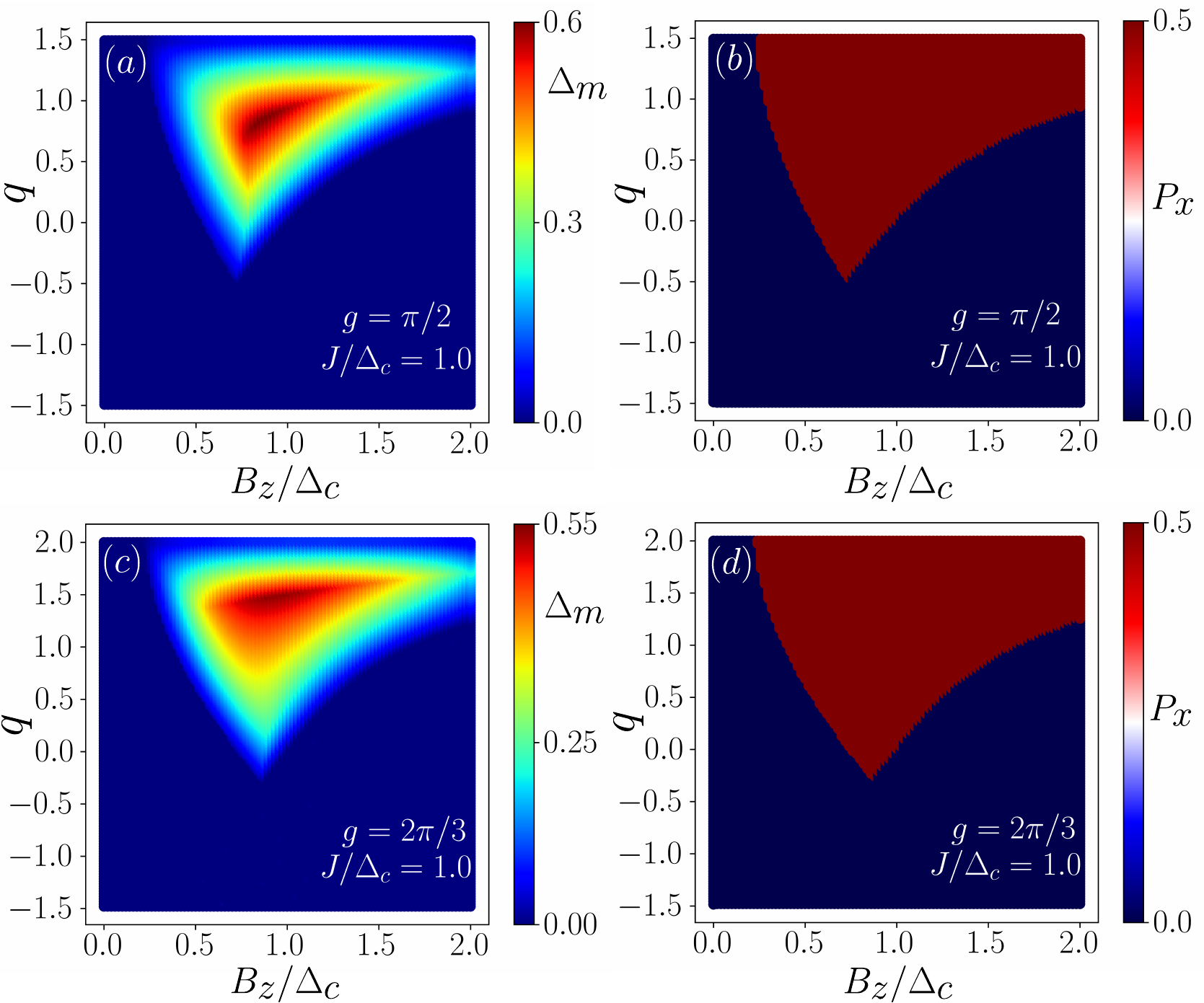}
\caption{\textbf{Topological phase diagram}:  In panels (a,b) and (c,d), the effective minigap $\Delta_m$ profile and the corresponding topological invariant: $P_x$ (bulk dipole moment) are depicted 
in the $B_z - q$ plane choosing $g=\pi/2$ and $g=2\pi/3$, respectively.  Other model parameters are chosen as $(t/\Delta_c, \mu/\Delta_c,J/\Delta_c, \Delta_c)=(0.5,1.0, 1.0 ,1.0~\text{meV})$.}
\label{fig:FigS4}
\end{figure}


\end{onecolumngrid}	
\end{document}